%% file: main.tex
\documentclass[sigplan,nonacm]{acmart}

\usepackage{listings}
\usepackage{algorithm}
\usepackage{algpseudocode}
\usepackage{subcaption}
\usepackage{diagbox}
\usepackage{multirow}
\usepackage{xcolor}
\usepackage{marginnote}
\usepackage{ifoddpage}
\usepackage{todonotes}

\AtBeginDocument{%
  }

\setcopyright{none}
\copyrightyear{2025}
\acmYear{2025}
\acmDOI{XXXXXXX.XXXXXXX}
\acmConference[Conference acronym 'XX]{Make sure to enter the correct
  conference title from your rights confirmation email}{June 03--05,
  2018}{Woodstock, NY}
\acmISBN{978-1-4503-XXXX-X/2018/06}

\begin{document}

\title{NeuroQD: A Learning-Based Simulation Framework For Quantum Dot Devices}

\author{Shize Che, Junyu Zhou, Seong Woo Oh, Jonathan Hess, Noah Johnson, Mridul Pushp, Robert Spivey, Anthony Sigillito, Gushu Li}
\affiliation{
  \institution{University of Pennsylvania}
 \country{USA}
}

\input{tex/0-Abstract}
\maketitle

\input{tex/1-Introduction}
\input{tex/2-Background}

\input{tex/3-Related-Work}

\input{tex/4-NeuroQD-Overview}

\input{tex/5-Observations}
\input{tex/6-Implementation}

\input{tex/7-Evaluation}

\bibliographystyle{ACM-Reference-Format}
\bibliography{sample-base}

\appendix
\input{tex/Appendix}

\end{document}

%% file: tex/0-Abstract.tex
\begin{abstract}
Electron spin qubits in quantum dot devices are promising for scalable quantum computing. However, architectural support is currently hindered by the lack of realistic and performant simulation methods for real devices. Physics-based tools are accurate yet too slow for simulating device behavior in real-time, while qualitative models miss layout and wafer heterostructure. We propose a new simulation approach capable of simulating real devices from the cold-start with real-time performance. Leveraging a key phenomenon observed in physics-based simulation, we train a compact convolutional neural network (CNN) to infer the qubit-layer electrostatic potential from gate voltages. Our GPU-accelerated inference delivers >1000× speedup with >96\% agreement to the physics-based simulation. Integrated into the experiment control stack, the simulator returns results with millisecond scale latency, reproduces key tuning features, and yields device behaviors and metrics consistent with measurements on devices operated at 9 mK.

\end{abstract}

%% file: tex/1-Introduction.tex
\section{Introduction}
\label{sec:introduction}

Electron spin qubits in semiconductor quantum dot (QD) devices are one of the leading candidates for building a large-scale quantum computer. Compared to other types of qubit, their distinct advantages include compatibility with mature CMOS fabrication in semiconductor infrastructure, long coherence time, fast gate operations, and small size for dense qubit integration~\cite{stuyck2024cmos, yoneda2018quantum, zajac2018resonantly, vandersypen2017interfacing}. Universal computing has been achieved for up to 6 qubits~\cite{philips2022universal, weinstein2023universal, zhang2025universal}, high single and two-qubit gate fidelities exceeding the fault tolerance threshold have been demonstrated~\cite{Xue, Noiri, Mills}, and industrial leaders are actively scaling this technology~\cite{george202412, eastoe2024method}, notably Intel, Quantum Motion, Diraq, Silicon Quantum Computing, 
and HRL~\cite{neyens2024probing, cai2019silicon, steinacker2025bell, thorvaldson2025grover, ha2021flexible}.

Although the use of electron spins in QDs as qubits was proposed long ago~\cite{DiVincenzo}, only recently have experimentalists developed device architectures capable of stable operation~\cite{veldhorst2015two, zajac2018resonantly, watson2018programmable, hendrickx2020fast}. This has positioned QD devices as an emerging quantum computing platform, for which architecture research remains scarce~\cite{chatterjee2021semiconductor}.

\subsection{Lack of Architectural Research and Suitable Simulation Tool}
A major concern for architecture research is how to design classical control systems that operate qubit-hosting devices in a programmable and scalable manner~\cite{jones2012layered}. However, the unique control challenges of QD devices make existing control architectures developed primarily for superconducting qubits inapplicable~\cite{fu2017experimental, stefanazzi2022qick, xu2023qubic, yang2022fpga, liu2025risc}. As a result, laboratories and companies build and test their own bespoke control infrastructure~\cite{khammassi2022scalable}, but in the absence of suitable simulation tools, the testing methods remain primitive (e.g. visually inspecting voltages on an oscilloscope).

Another research area unique to QD devices and critical to architecture design is tuning software~\cite{baart2016computer, mcjunkin2021heterostructure, van2022all, darulova2020autonomous, lapointe2020algorithm, czischek2021miniaturizing, Kalantre2019, zwolak2020autotuning, van2022all, ziegler2023automated, che2024fast, hickie2024automated, rao2025modular}. These programs execute procedures to bring the device into an operational state and are essential for the scalability of QD devices~\cite{zwolak2023colloquium}. However, developing and testing these softwares without access to actual devices remains difficult due to the same lack of suitable simulators. A recent review on QD device tuning points out the need for simulation to cover all aspects of the tuning process~\cite{zwolak2024data}.

In general, testing control hardware/software on devices is costly because QD devices are fragile and can be permanently damaged by control bugs~\cite{dodson2020fabrication}. In classical computer architecture, researchers routinely develop and validate ideas on simulators without fabricating chips~\cite{binkert2011gem5, sanchez2013zsim, kim2015ramulator, bakhoda2009analyzing, khairy2020accel}. Applying the same idea to QD devices by developing simulators to interface \emph{in~situ} with control systems is sensible given the high cost and risk.

\subsection{Criteria for Architectural Simulation of QD}

Simulations serve different purposes and vary in how closely they model physical reality. For \emph{in~situ} interfacing with control systems, we identify two essential criteria. First, sufficient physical-level accuracy to capture the behavior of the target device. Second, real-time response to match the expectations of the control system. These criteria are inherently in tension because higher physical-level accuracy incurs greater computational cost and latency. The goal is to develop a simulation framework that balances physical accuracy and performance for practical, control-in-the-loop use.

\begin{figure}[t]
  \centering
\includegraphics[width=0.8\columnwidth]{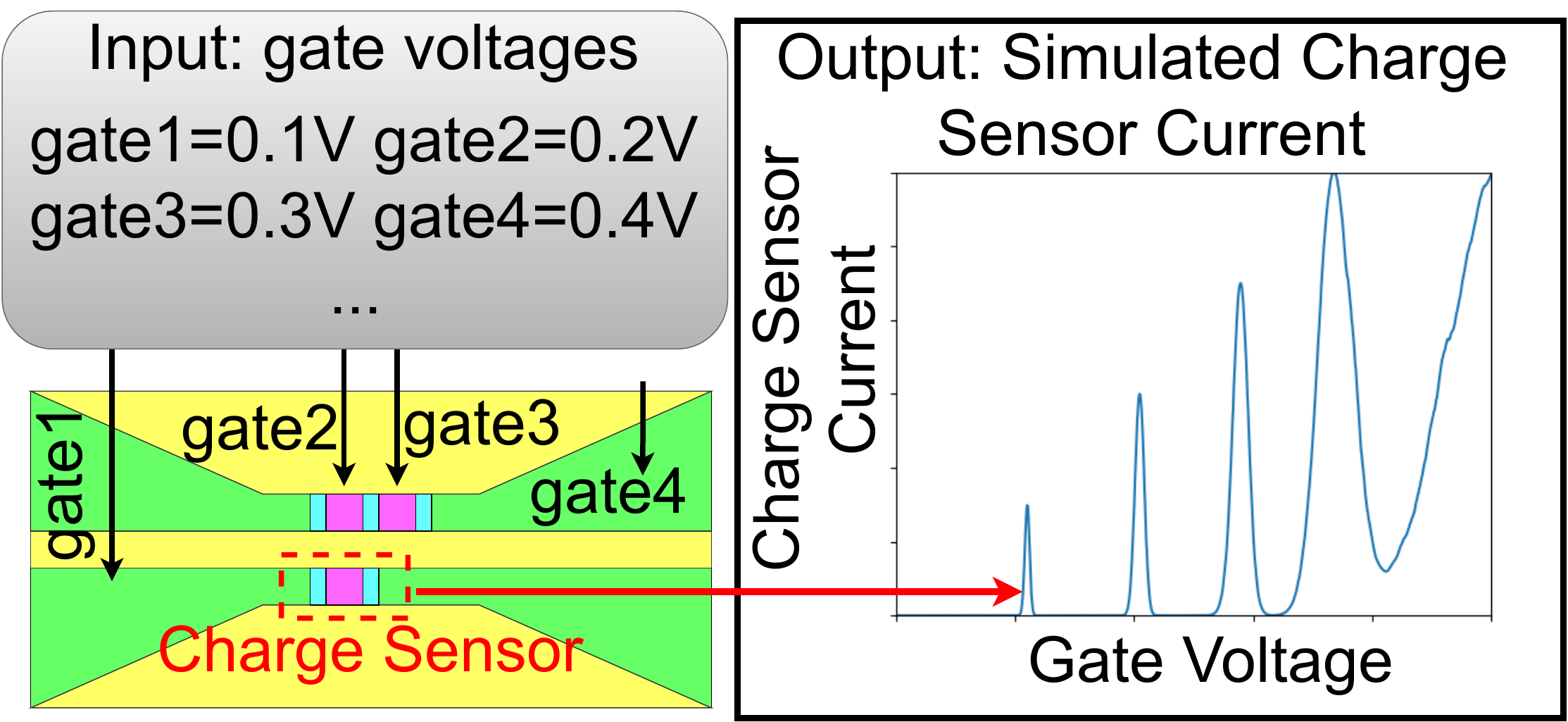}\vspace{-5pt}
  \caption{Overview of the target quantum dot device simulation}\vspace{-20pt}
  \label{fig:device-simulator}
\end{figure}

Different from other works~\cite{johansson2012qutip, qiskit2024, The_CUDA-Q_development_team_CUDA-Q, jiang2025bqsim} on quantum program or Hamiltonian simulation, this paper targets device-level simulation of QD devices and its charge-sensor response to applied gate voltages. As illustrated in Figure \ref{fig:device-simulator}, the target simulator takes as input the voltages applied to the device’s gate electrodes and produces as output the charge sensor current. Note that in this paper, the term `gate' refers to the physical gate electrodes in hardware, not the unitary transformation in software.

Current simulation methods tend to optimize for one side of the trade-off: either achieving high physical accuracy at the expense of latency (\textbf{Type 1}), or running fast but with oversimplified device models (\textbf{Type 2}).

\subsection{Overview of NeuroQD}
In this paper, our objective is to develop a fast yet realistic simulator that meets both the accuracy and performance criteria. To ensure physical integrity, our methods will be rooted in the \textbf{Type 1} simulation paradigm while we made two observations about the time-consuming physics-based simulation.
\textbf{First,} the simulation does not have to be accurate for all possible gate voltage configurations. In practice, all the gate voltages are limited in some specific ranges and the simulation only needs to work in these ranges.
\textbf{Second,} calculating the 2-dimensional electron gas (2DEG) potential, which is the performance bottleneck in the \textbf{Type 1} simulation workflow, can be regarded as a fully-convolutional blurring transformation from the input gate voltage configuration.
Both of them suggest that a machine-learning-based simulation can be the key to bypassing the performance bottleneck in \textbf{Type 1} physics-based simulation.

To this end, we propose NeuroQD, an end-to-end machine-learning (ML)-based simulation framework for QD devices that can achieve both \textbf{Type 1}-level accuracy and real-time performance.
\textbf{First,} to collect the training data for our ML model, we employ the physics-based simulation in COMSOL, a commercial electromagnetic solver widely used for QD device simulation~\cite{zajac2016scalable, zajac2015reconfigurable, sokolov2020simulation, liang2024electronic, ma2020three, hardy2019single, tadokoro2021designs, mcjunkin2021valley, cifuentes2023path, cifuentes2024bounds, miller2022effective, hong2021developing, duan2020remote}, to generate the 2DEG training data. 
To overcome the difficulty of sampling from an extremely large input space, we propose a new graph-based sampling method together with Latin Hypercube sampling and data augmentation to identify critical subspaces and reduce the time required to generate training data from the slow physics-based simulation.
\textbf{Second,} since the 2DEG potential calculation can be considered as a fully-convolutional transformation, we propose to use a convolutional neural network (CNN) architecture and select the U-Net model, which is well-suited for learning the target blurring transformation due to its capability of capture both large and fine-grained spatial details via the encoder-decoder architecture with skip connections.
\textbf{Third,} we develop a complete end-to-end simulator pipeline by fulfilling the physics-based post-processing with multiple algorithms and integrating the simulator into our in-house QD device experimental infrastructure. 

We conducted comprehensive evaluation on our proposed simulation pipeline.
Our simulation framework generates the training data from COMSOL and train the U-Net on 2-dot device only. Driven by the translation invariant and input-size agnostic nature of CNN, the trained model can naturally generalize the 2DEG potential inference beyond the 2-dot device it is trained on, which justifies the training cost and validates the scalability of our method. Comparing to the COMSOL baseline on devices spanning 2-dot to 12-dot, our simulator can achieve over 1,000× speedup while maintaining >96\% accuracy and a 100\% success rate compared to COMSOL’s 91.33\%. We further tested the simulation from 12-dot to 99-dot devices and our results show that accuracy remains high with the runtime grows linearly with respect to the number of dots, suggesting scalability beyond the state-of-the-art devices.

We compare the performance and simulated data of a 2-dot integrated simulation backend to two real devices: a 2-dot device and the left 2-dot of a 4-dot device. Specifically, we run four experiments from cold-start to charge state tuning: turn-on, slit vs sensor sweep, Coulomb peaks, and charge stability. Features in data taken on real devices are accurately reconstructed in the simulation backend, and the important physical quantities inferred from the simulated data are comparable to the experimental data from both real devices. The performance of the simulator is as fast as live measurements on devices.

Our major contributions can be summarized as follows:
\begin{enumerate}
    \item We propose an ML-based simulation framework that can simulate the behavior of QD devices quickly and accurately.
    \item We made several observations and improve the ML model training with optimizations in training data sampling. 
    \item Experimental results show that our ML model can be generalized to different QD devices, and achieve significant speedup with high accuracy compared with the expensive physics-based simulation baseline.
    \item We integrate the proposed simulator into our experimental infrastructure and find that the simulated results match well to real device experiments.
\end{enumerate}

%% file: tex/2-Background.tex
\section{Background}
\label{sec:background}

\subsection{Quantum Dot Devices}
\label{sec:quantum-dot-devices}
Quantum dot devices are semiconductor devices that can trap and control individual electrons to encode qubits~\cite{DiVincenzo, hanson2007spins, burkard2023semiconductor, vandersypen2019quantum, hu2025single}. The trapping of individual electrons is achieved by providing confinement potential in both the vertical direction and the lateral direction: vertically, through a layered semiconductor heterostructure that forms a quantum well; and laterally, by applying voltages to gate electrodes to shape the potential in the 2DEG layer.

\subsubsection{Heterostructure}
\label{sec:heterostructure}
\begin{figure}[t]
  \centering
  \includegraphics[width=\columnwidth]{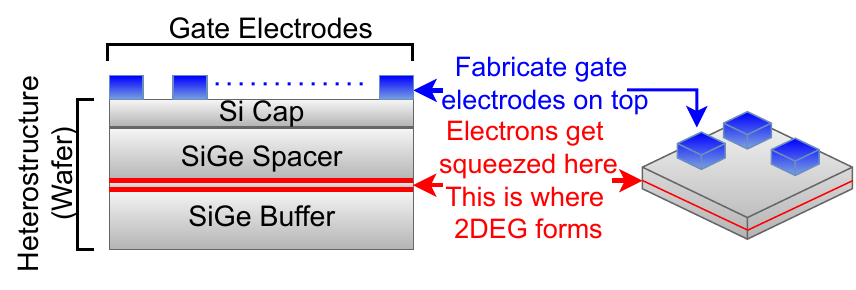}
  \caption{Schematic of Si/SiGe heterostructure. Gates are fabricated on top and the electrons are trapped in the middle layer.}
  \label{fig:heterostructure}
\end{figure}

We begin by examining the heterostructure, which refers to the layered composition of the wafer used in device fabrication. As illustrated on the left side of Figure~\ref{fig:heterostructure} (a cross-sectional view of the device), the wafer is composed of several distinct layers: a Si cap, a SiGe spacer, a central layer highlighted in red, and a SiGe buffer. These layers are made of carefully chosen semiconductor materials designed to create a natural vertical confinement potential. This confinement squeezes electrons into a narrow region defined by the two red lines in the center of the stack. This region, known as the quantum well, effectively traps the electron wavefunction along the vertical direction. As a result, a very thin layer of electrons forms at a specific depth below the surface, commonly referred to as the two-dimensional electron gas (2DEG) layer.

\subsubsection{Gate Electrodes}
\label{sec:gate-electrodes}
While the heterostructure provides vertical confinement, lateral confinement is achieved by fabricating gate electrodes on top of the wafer, as illustrated by the blue rectangles in Figure~\ref{fig:heterostructure}. By applying carefully tuned voltages to these gates, the electrostatic potential in the 2DEG layer can be precisely shaped. This allows for the isolation of individual electrons in defined regions of the 2DEG, enabling the use of their spin states to encode qubits.

\subsubsection{Device Operation and Tuning}
\label{sec:device-operation-and-tuning}
Devices are operated by the control system, which is a large collection of hardware, firmware, and software that performs mainly two operations to tune the device into an operational state: \textbf{set} gate voltages, and \textbf{read} charge sensor current. 

\begin{figure}[t]
  \centering
  \includegraphics[width=\columnwidth]{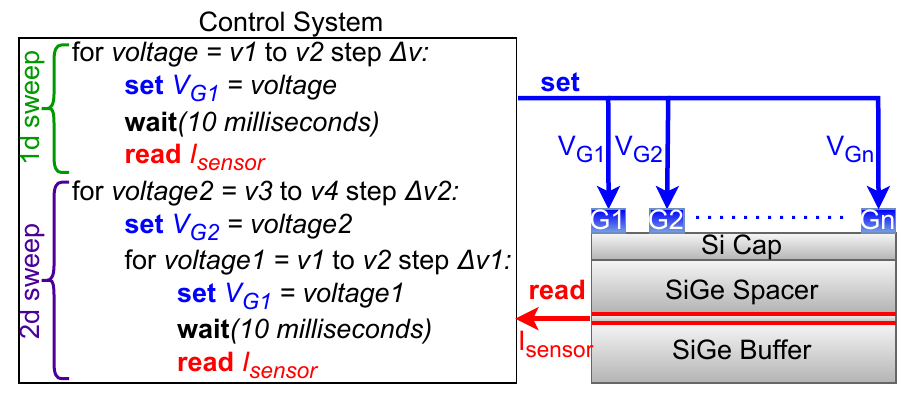}
  \caption{An example of interaction between control system and device}
  \label{fig:op-and-tuning}
\end{figure}

Typically, the interactions between the control software and the device are numerous 1D and 2D voltage sweeps composed by looped \textbf{set} and \textbf{read} operations for device tuning up. As illustrated in Figure~\ref{fig:op-and-tuning}, a sweep involves iterating over a sequence of voltage values, typically 50 to 1000 points depending on the step size $\Delta v$, applied to one or two gates (G1 and G2 in Figure~\ref{fig:op-and-tuning}). For each gate voltage configuration, the system executes a \textbf{set}-\textbf{wait}-\textbf{read} sequence: it sets the gate voltages, waits for a short period to allow the circuit to settle, and then reads out the charge sensor current. The \textbf{wait} step, which typically lasts from a few microseconds to tens of milliseconds, compensates for signal delays introduced by heavy filtering, dividing, and amplifying in the analog signal path. By sweeping over various gate combinations and recording the charge sensor current, the system generates a collection of 1D and 2D plots that reveal key device behaviors. These plots guide the control software to identify the appropriate operating voltage ranges for each gate.

Due to this control system operating scheme, our target simulator needs to take as input the set of gate voltages applied by the control system, and calculate the charge sensor current within the millisecond \textbf{wait} window. This is the real-time performance requirement that we seek to achieve in this work. In addition to the performance criteria, we seek to achieve high physical-realism such that the 1D and 2D plots taken during device tuning can be simulated.

%% file: tex/3-Related-Work.tex
\section{Related Work}

\textbf{Existing QD device simulation tools and research.} Existing simulation tools for QD devices can be classified into two categories: \textbf{Type 1:} \textit{First-principle physics-based simulations}, \textbf{Type 2:} \textit{Fast, qualitative simulations} based on the constant interaction model (CIM). 

\textbf{Type 1} simulations generally use first-principle approach where exact device physical architecture is created using a CAD interface and the electrostatics is solved self-consistently using finite-element analysis and a set of differential equations (e.g. Poisson equations). These simulations are typically embedded as part of commercial scientific software, and notable examples that have been used for QD device simulation include COMSOL~\cite{multiphysics1998introduction}, TCAD~\cite{ikegami2019tcad, asai2021development}, nextnano~\cite{birner2007nextnano}, QCAD~\cite{gao2013quantum}, QTCAD~\cite{beaudoin2022robust}, MaSQE. These simulation tools are close to ground truth but computationally expensive, as such, they are widely used for providing the 2DEG potential for theoretical studies on the exact quantum mechanical physics in 2DEG~\cite{anderson2022high, cifuentes2023path, cifuentes2024bounds}. 

\textbf{Type 2} simulations are typically open source and implement CIM, examples include QDSim~\cite{gualtieri2025qdsim}, QDarts~\cite{krzywda2025qdarts}, QArrays~\cite{van2024qarray}. CIM is a simplified QD device model where the coupling between dots and gates are treated as capacitors with assumed capacitance and the quantum dot device is then modeled as a network of capacitors~\cite{van2002electron, hanson2007spins, kouwenhoven1997electron, schroer2007electrostatically}. While fast for near-term devices, this framework ignores detailed device architecture and can only simulate charge stability diagrams, as such, it is incomprehensive for interfacing with control system where other hardware-observable behaviors also need to be covered. Furthermore, CIM’s energy minimization is NP-hard, limiting scalability to larger devices.

Other more focused qualitative models exist such as SimCATS~\cite{hader2024simulation} which generates CSDs geometrically, but generally suffer the same problem of ignoring device details and not covering enough hardware-observable behaviors.

Our method uses COMSOL's electrostatics simulation as the baseline~\cite{multiphysics1998introduction}, which is one of the most accurate way to calculate the 2DEG potential from exact device physical architecture~\cite{cifuentes2023path} and has served as the standard tool for simulating realistic QD devices both theoretically and experimentally~\cite{hardy2019single, tadokoro2021designs, mcjunkin2021valley, cifuentes2023path, cifuentes2024bounds, miller2022effective, hong2021developing, duan2020remote}. While theory-focused works typically put the potential solved by COMSOL into multi-electron Hamiltonian to derive fine electron behavior, we do not take this path as it incurs the classic "using classical computer to simulate quantum" problem to reach a physical-level detail beyond the requirement for our practical purpose at a serious performance penalty. Instead, we adopt a mesoscopic electron density model (Section~\ref{sec:physics-model} and Section~\ref{sec:charge-state-model}) that allows fast but relatively coarse calculation of the charge state. Prior works have confirmed the applicability of this model in COMSOL by comparing it to real devices~\cite{zajac2015reconfigurable, zajac2016scalable, ZajacThesis}. 

All the COMSOL procedures used in this work are automated using COMSOL's Java API. The electrostatics module in the latest version of COMSOL 6.3 does not support GPU acceleration~\cite{COMSOL63}, so we made best effort to ensure maximum parallelism by executing all the COMSOL workloads using all 24 cores on the CPU. 

\textbf{Device architectures.}
In this work, we focus on simulating our in-house fabricated devices which adopt the Si/SiGe dot array architecture introduced in Section~\ref{sec:background}. We focus on these devices as we have the complete source information from gate design to heterosteucture, which allows faithful modeling of these devices in COMSOL and compare the simulation to experiments. While gate architectures that implement $2\times n$ and $n \times n$ dot arrays were demonstrated (more common on Ge/SiGe and SiMOS heterostructrues)~\cite{unseld20232d, gilbert2020single, hendrickx2021four, zhang2025universal}, $1 \times n$ dot array remains the most mature device architecture as of today~\cite{philips2022universal}. 


%% file: tex/4-NeuroQD-Overview.tex
\section{Overall Approach}
The overall approach taken by NeuroQD can be described as identifying the performance bottleneck of the COMSOL baseline and training an accurate surrogate model to replace it to meet the performance requirement. Surrogate modeling using machine learning has been applied to a wide range of scientific problems~\cite{guan2023using, lam2023learning, galletti20255d, chen2022high, spurio2022cosmopower}. In our case, we apply surrogate modeling to the 2DEG potential solving step (bottleneck) in the COMSOL baseline.

\begin{figure}[t]
  \centering
  \includegraphics[width=0.95\columnwidth]{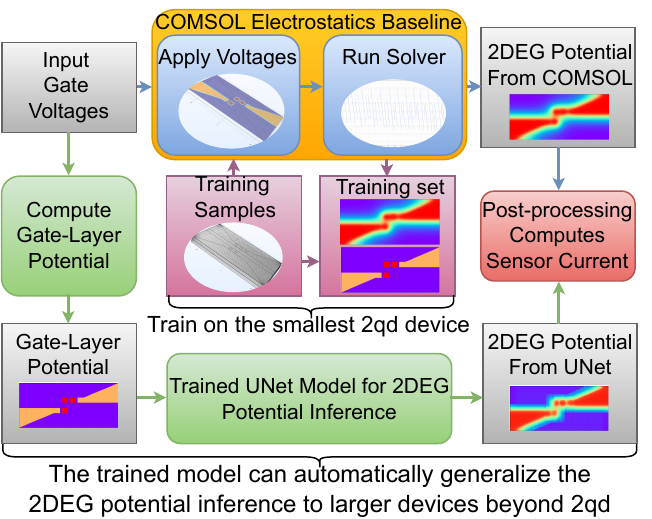}
  \caption{Overall approach}
  \label{fig:overview}
\end{figure}

Figure~\ref{fig:overview} demonstrates the overall approach. The blue arrows represent the datapath of the baseline COMSOL simulation. Specifically, it takes as input the gate voltages. The "Apply Voltages" step encodes gate voltages as a set of boundary conditions into the system of differential equations. The "Run Solver" step solves the PDEs self-consistently, which produces the 2DEG potential. The post-processing steps calculate the charge sensor current from the 2DEG potential. In this baseline simulation datapath, the performance bottleneck is solving the PDEs to produce the 2DEG potential. The self-consistent solving is iterative in nature and typically takes longer than 10s, compared to the milliseconds latency we aim for. In addition, self-consistent solving is not guaranteed to converge, which makes it unable to produce results for some gate voltage configurations. A surrogate model would solve both the performance and the convergence problem.

The challenge in training a surrogate model in place of the "Apply Voltage" and "Run Solver" steps lies in 1) model complexity and 2) scalability. A brute-force approach would replace these two steps with a black-box model that maps gate voltages directly to the 2DEG potential. This approach inevitably requires high model complexity, as it must emulate the entire PDE solving. This approach is also not scalable, as a network trained on a small device does not readily transfer to larger ones, forcing device-specific retraining.

Unlike the brute-force approach, our approach relies on two important observations on the COMSOL electrostatics simulation baseline that enabled us to achieve 1) high accuracy with a compact U-Net and 2) train on the smallest device and generalize to larger devices. The green arrows in Figure~\ref{fig:overview} represent the datapath of our surrogate approach. First, we calculate the gate-layer potential (easily computed from gate layout and gate voltages), then a trained U-Net model maps the gate-layer potential to the 2DEG potential. The mapping from gate-layer potential to 2DEG potential naturally generalizes to larger devices.

In Section~\ref{sec:observations}, we detail the two observations. In Section~\ref{sec:simulator-pipeline}, we detail the implementation, which focuses on physics models (Section~\ref{sec:physics-model}) and sampling optimizations (Section~\ref{sec:sampling-method}) for training data generation, as well as integration of the final simulator into our experiment software. Note that our key technical focus is on establishing a new framework to efficiently solve the simulation problem, rather than optimizing the ML techniques. Overall, our approach has made it possible to achieve high accuracy on a highly simple model trained on the smallest device, so further ML optimizations do not make sense and are not our main focus.

%% file: tex/5-Observations.tex
\section{Observations}
\label{sec:observations}

\subsection{Observation 1: Bounded Voltage Space}
\label{sec:observation-1}
Our first observation is based on an experimental constraint. QD devices operate reliably only within a strict gate voltage tolerance range. Applying voltages outside this range can irreversibly damage the device. For our in-house-fabricated Si/SiGe devices, this safe operating range is 0–1V. In practice, we enforce this constraint through multiple layers of software assertions to ensure that no out-of-bounds voltages are ever applied during an experiment.

This constraint has an important implication for simulation: the effective space of valid gate voltage configurations is fixed and known. Therefore, a simulator does not need to generalize beyond this bounded voltage domain, allowing models to focus their capacity on a small and well-defined input space.

\begin{figure}[t]
  \centering
  \includegraphics[width=0.95\columnwidth]{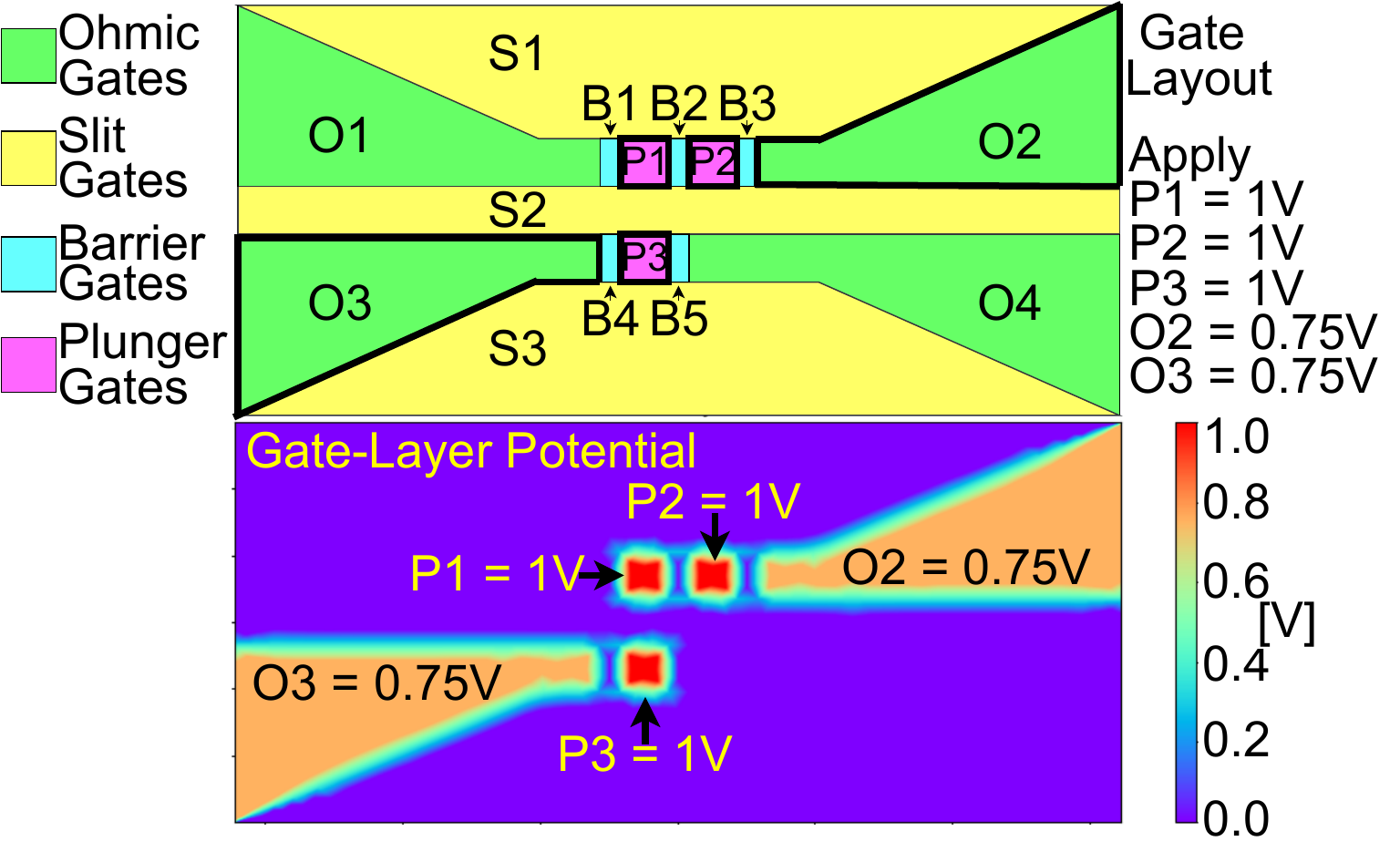}
  \caption{Gate-layer potential from gate layout}
  \label{fig:gate-layer-potential}
\end{figure}

\subsection{From Gate Voltages to 2DEG Potential}
\label{sec:from-gate-voltages-to-2deg-potential}
Before introducing our second observation, we first describe the physical transformation from applied gate voltages to the resulting potential in the 2DEG layer.

In the gate layer (the top surface of the wafer), each gate defines an electrostatic potential in the region directly beneath it. As a result, the gate-layer potential is partitioned into disjoint regions, each governed by the voltage applied to a specific gate.

Figure~\ref{fig:gate-layer-potential} illustrates this effect. The top panel shows the gate layout of the 2-dot device, where each region is labeled with its corresponding gate name and colored according to gate type. For example, the three pink regions correspond to the plunger gates P1, P2, and P3, while the four green regions represent the ohmic gates O1, O2, O3, and O4. When a specific voltage configuration is applied, for instance:
\[
\{\text{P1}=1V, \text{P2}=1V, \text{P3}=1V, \text{O2}=0.75V, \text{O3}=0.75V\}
\]
with all other gate voltages set to 0V, the resulting gate-layer potential is shown directly below the layout in Figure~\ref{fig:gate-layer-potential}. As shown, each activated gate region maintains a nearly uniform potential equal to its assigned voltage, with narrow transition bands at the edges due to physical boundary smoothing.

This potential profile is defined at depth = 0 nm, corresponding to the gate layer, and we name it gate-layer potential. As we move deeper into the wafer, each layer has a distinct potential shaped by the electrostatics of the heterostructure. The 2DEG layer, where quantum dots reside, is located 59 nm below the surface. The transformation from the gate-layer potential to the 2DEG potential can thus be understood as a physical "propagation" process through the wafer heterostructure.

\begin{figure*}[t]
  \centering
  \includegraphics[width=0.9\textwidth]{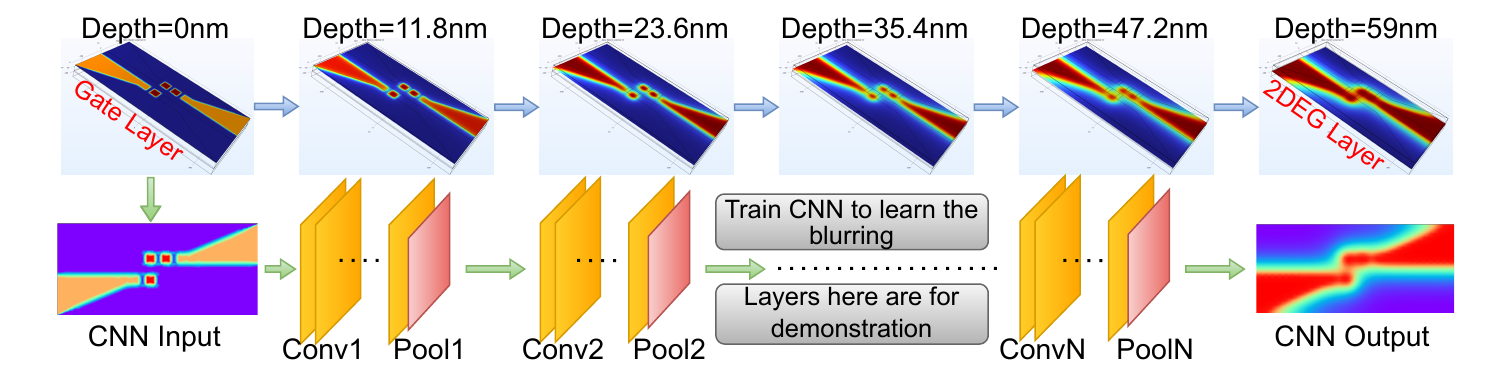}
  \caption{Potential profile at different depths as it traverses down the dielectrics. We model this blurring effect using CNN.}
  \label{fig:potential-blurring}
\end{figure*}

\subsection{Observation 2: High-Order Blurring Transformation}
\label{sec:observation-2}
To better understand this transformation, we examine the potential profile at several depths: 0 nm  (the gate layer), 11.8 nm, 23.6 nm, 35.4 nm, 47.2 nm, and finally 59 nm (the 2DEG layer), using the same gate voltages as above. Figure~\ref{fig:potential-blurring} plots these potentials. The top row of Figure~\ref{fig:potential-blurring} shows how the initially sharp gate-defined boundaries gradually blur and diffuse with depth. By the time the potential reaches the 2DEG layer, the original structure has been smoothed into a continuous, soft-edged profile.
This transformation resembles a high-order blurring operation, a fully-convolutional transformation well suited to be modeled by a CNN.

Based on this insight, we propose replacing the physics-based computation of the 2DEG potential with a CNN trained to approximate this transformation. A CNN trained on data generated from COMSOL can act a surrogate model for the slowest step in the baseline simulation datapath. This substitution addresses the performance and convergence challenges in the COMSOL baseline:
(1) The CNN input--the gate-layer potential--can be easily generated by assigning gate voltages to their respective gate regions. 
(2) CNN inference is highly parallelizable and can be accelerated by modern GPUs, enabling real-time performance. 
(3) CNN always produces a valid output, avoiding the convergence failures encountered in COMSOL's iterative solver. 
This modeling approach forms the foundation of our real-time simulation framework, offering both physical plausibility and orders of magnitude performance improvement.

The most significant advantage of learning this transformation is generalization to larger devices. The transformation from gate-layer to 2DEG potential is defined by the heterostructure and is independent of gate-layout. Therefore, a CNN that learns this transformation, because it is translationally invariant and input-size agnostic~\cite{long2015fully}, can be readily applied to devices with the same heterostructure but larger gate layout. As we demonstrate in the evaluation section, a model trained exclusively on a 2-dot device can retain high accuracy when applied to devices with up to 99 quantum dots.

%% file: tex/6-Implementation.tex
\section{NeuroQD Implementation}
\label{sec:simulator-pipeline}

In this section, we detail the implementation of the proposed simulation framework built directly on the two core observations introduced. We begin by capturing the gate layout and wafer structure of our in-house fabricated Si/SiGe devices and construct a physics-based model in COMSOL to generate training dataset on 2-dot device (2qd). The U-Net model~\cite{ronneberger2015u} is then trained to learn the transformation from the gate-layer potential to the 2DEG potential. Once trained, the model serves as a fast, learned approximation of the COMSOL simulation step, which eliminates the computational bottleneck in the baseline. Finally, we apply a lightweight post-processing pipeline to convert the predicted 2DEG potential into simulated device responses, and integrate the full simulator into our experiment control software as a real-time-backend to interface with the \textbf{set} and \textbf{read} APIs.

\subsection{Generating Data From COMSOL}
\label{sec:generating-data-from-comsol}

\begin{figure}[t]
  \centering
  \includegraphics[width=0.84\columnwidth]{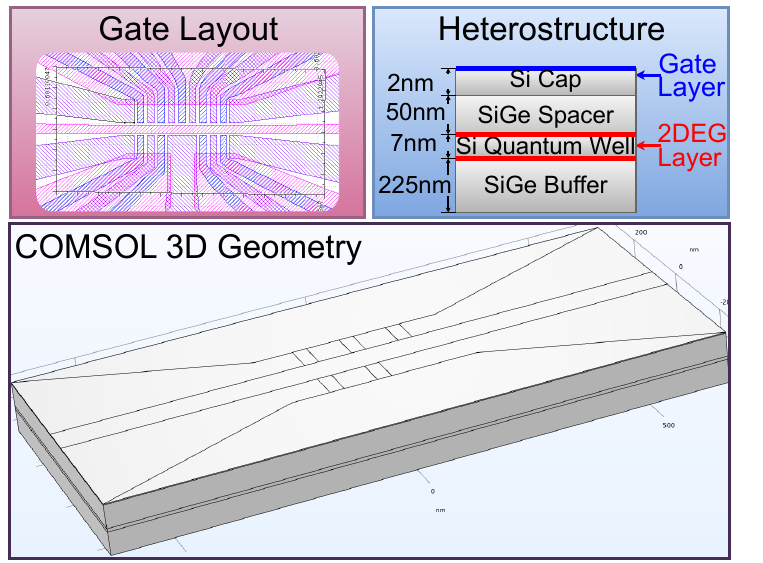}
  \caption{Device geometry based on gate layout and heterostructure}
  \label{fig:device-geometry}
\end{figure}

We generate 2DEG potentials using the COMSOL electrostatics module, with simulation models grounded in the physical design of our in-house fabricated Si/SiGe devices. Specifically, we construct full 3D geometries of the 2-dot device that accurately capture both the gate dimensions and the heterostructure of the wafer. To model the electrostatics at the 2DEG layer, we incorporate a physics model for the ideal 2DEG, which defines the relationship between electrostatic potential and charge density. We propose a graph-based sampling method combined with Latin Hypercube sampling to sample the gate voltage configurations used for training.


\subsubsection{Device 3D Geometry}
\label{sec:device-3d-geometry}

We capture gate dimensions and approximate gate layout using the design of a 4-dot device to construct the 3D geometry of the 2-dot device used for training in COMSOL. The upper left panel of Figure~\ref{fig:device-geometry} shows the gate layout, with the simulation region defined by the black rectangle. This rectangular region defines the shape and dimensions of the gates. To build the 2-dot device, we shrink the number of quantum dots while keeping the gate dimensions fixed. The lower panel of Figure~\ref{fig:device-geometry} shows the 3D model of the 2-dot used for training. 



In addition to gate layout, the 3D model captures the vertical heterostructure of the wafer. The upper right panel of Figure~\ref{fig:device-geometry} shows a schematic cross section of the wafer, which we replicate in the COMSOL geometry. From top to bottom, the device consists of four layers: 2nm Si, 50nm Si$_{0.7}$Ge$_{0.3}$, 7nm Si (where the 2DEG forms), and 50nm Si$_{0.7}$Ge$_{0.3}$. While the actual SiGe buffer layer is thicker (225nm), we use 50nm in simulation, as the thickness of layers below 2DEG does not affect the results.

\subsubsection{Physics Model}
\label{sec:physics-model}
We use ideal 2DEG physics to model the relation between electron density and potential in the 2DEG layer. The carrier density in the first subband (electrons in semiconductors occupy the first subband mostly at very low temperature) of an ideal 2D electron system is given by~\cite{jena2022quantum}:
 $ n_{2D} = \frac{m^* k_B T}{\pi \hbar^2}ln(1 + exp({\frac{E_F - E_1}{k_BT}}))$
where $m^*$ is the effective electron mass in Silicon, $k_B$ is the Boltzmann constant, $T$ is the temperature, $\hbar$ is the reduced Planck constant, $E_F$ is the Fermi energy, and $E_1$ is the ground state energy. In the case of 2DEG in quantum dot devices, the Fermi energy $E_F$ is the 2DEG turn-on potential $V_{on}$ times the single electron charge $-e$ ($e$ is the electrary charge), where the turn-on potential $V_{on}$ remains constant, and the ground state energy is the 2DEG potential $V$ times the single electron charge $q_e$. With $E_F = -eV_{on}$, $E_1 = -eV$, we can convert the formula into a function of the 2DEG potential $V$ where
  $n_{2D}(V) = \frac{m^* k_B T}{\pi \hbar^2}ln(1 + exp({\frac{e(V - V_{on})}{k_BT}}))$.

This function specifies the relation between the 2DEG potential and the electron density, but it is relatively complicated and takes a significant time for the solver to converge. Therefore, we further simplify it by assuming an ideal temperature of $0K$ ($T = 0K$). When $T = 0K$, the $exp({\frac{e(V - V_{on})}{k_BT}})$ term dominates inside the natural logarithm. Therefore, we have
\begin{align}
\begin{aligned}
  n_{2D}(V) &= \frac{m^* k_B T}{\pi \hbar^2}ln(1 + exp({\frac{e(V - V_{on})}{k_BT}}))\\
  &\overset{T = 0K}{\approx} \frac{m^* k_B T}{\pi \hbar^2}ln(exp({\frac{e(V - V_{on})}{k_BT}}))
   =  \frac{m^* e(V - V_{on})}{\pi \hbar^2}
\end{aligned}
\end{align}
In this simplified form, when $V < V_{on}$, the electron density $n_{2D}$ (unit $1/m^2$) takes negative values, which makes no sense. On the other hand, the case $V < V_{on}$ is valid in the original function with the electron density $n_{2D}$ exponentially decaying to 0. Therefore, we add a piecewise $n_{2D} = 0$ for the case $V < V_{on}$. The final form is then
\[
n_{2D}(V) =
\begin{cases} 
    \frac{m^* e(V - V_{on})}{\pi \hbar^2}, & V \geq V_{on} \\
    0, & V < V_{on}
\end{cases}
\]
which is our physics model for simulating the relation between electron density and potential at the 2DEG layer. This relation is encoded as part of the COMSOL simulation to solve the 2DEG potential.

\subsubsection{Sampling Method}
\label{sec:sampling-method}

We generate the training set using a combination of two sampling strategies: Latin Hypercube Sampling (LHS) and a custom graph-based method.


\textbf{Difficulty of Training Data Generation:} While the observations allow training on the smallest 2-dot device with a fixed gate voltage range, the sampling space is still large. The 2-dot device has 15 gates; Assuming 1mV voltage resolution, there are a total of $10^{45}$ possible gate voltage configurations. The sampling should cover this space as much as possible and cover important subspaces that matter to the overall simulation quality. 

Because we consider the target transformation a fully-convolutional blurring that blurs the boundary between gate voltage regions, the important subspaces are the connected gates from which the local interactions can be learned. For example, using Figure~\ref{fig:gate-layer-potential} as a reference, P1 and S1 are connected, so their boundary will become blurred by the transformation, affecting their respective regions. We therefore consider the subspace spanned by P1 and S1 important, as these two regions have local interactions that need to be learned. In contrast, gates P1 and S3 are not connected, so their voltages will have little effect on their respective regions. We therefore consider the subspace spanned by P1 and S3 to be unimportant because they do not have local interactions in the transformation. In the context of QD devices, this local interaction corresponds to the cross-coupling between the gates~\cite{hensgens2017quantum}, which means that the potential in a region that should be controlled by a particular gate is also affected by nearby gates. As cross-coupling is responsible for many phenomena observed during device tuning, the sampling method needs to identify and cover cross-coupled subspaces.

We apply Latin Hypercube Sampling (LHS) on the entire voltage space and use a graph-based method to identify and sample subspaces with strong cross-coupling. Given a total training sample budget of \( n \), we allocate half of the samples to LHS and half to graph-based sampling.

    \textbf{Latin Hypercube Sampling (LHS):} 
    LHS achieves broad and uniform coverage of the 15-dimensional gate voltage space~\cite{mckay2000comparison, stein1987large}. This helps the model learn the general structure of the gate-to-2DEG transformation across a wide range of configurations.

\begin{figure}[t]
  \centering
  \includegraphics[width=0.8\columnwidth]{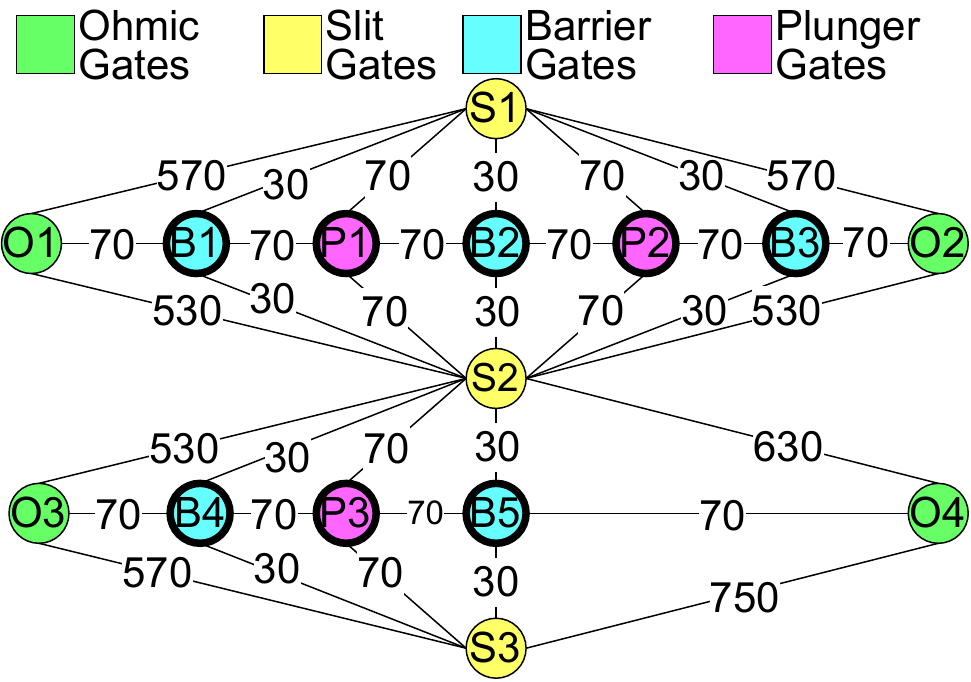}
  \caption{Graph for extracting cross-coupled subspaces}
  \label{fig:2qd-graph}
\end{figure}

    \textbf{Graph-Based Sampling:} 
    We define a weighted undirected graph.
        Each vertex represents a gate in the 2qd layout.
        An edge exists between two gates if their gate regions are connected.
        Edge weights correspond to overlap distances and serve as a proxy for cross-coupling.
    This graph (Figure~\ref{fig:2qd-graph}) captures all connected sets of gates where cross-couplings exist and uses the overlap distance as a way to quantify the magnitude. The sampling focuses on the bold nodes as these gates cover the critical dot region where electrons are confined. We extract all 2-node and 3-node subgraphs that include bold nodes, which are all the 2-gate and 3-gate cross-coupled subspaces that affect the dot region.
    For each subgraph with total edge weight \( W_{\text{sub}} \), we assign  $\frac{W_{\text{sub}}}{W_{\text{total}}} \times \frac{n}{2}$ samples, where \( W_{\text{total}} \) is the total edge weight of all subgraphs. For each subgraph, we apply LHS within its corresponding subspace while keeping all other gate voltages at zero. This graph-based sampling method ensures that the training set contains samples enriched with key local interactions.

Together, these two strategies produce a training set that balances global coverage with localized detail, enabling the CNN to learn both broad transformations and subtle coupling effects.

\subsection{U-Net Model}
\label{sec:unet-model}

We train the U-Net model using the generated training set. We augment the data generated from COMSOL via rotation and reflection detailed in Section~\ref{sec:data-augmentation} in the appendix. A total of 20,000 samples are allocated for the generation of training set, and after running them through COMSOL, 18,442 samples converged. With the data augmentation applied, the final training set consists of 147,536 samples.
The architecture of the U-Net model is summarized in Table~\ref{unet-summary}. As optimizing the ML model is not the focus of this work, we keep this section short and put the details in the appendix.

\subsection{2DEG Potential Post-Processing}
\label{sec:2deg-potential-post-processing}
Once the U-Net model generates the 2DEG potential, additional post-processing is required to compute physical observables such as the charge state of each quantum dot (i.e., the number of confined electrons) and the sensor dot current. This section outlines a lightweight physics-based procedure for extracting these quantities.


\subsubsection{Charge State Model}
\label{sec:charge-state-model}

As described in Section 6.1.2, the electron density at the 2DEG layer is governed by the following relation:
$
n_{2D}(V) =
\begin{cases} 
    \frac{m^* e(V - V_{on})}{\pi \hbar^2}, & V \geq V_{on} \\
    0, & V < V_{on}
\end{cases}
$. 
To determine the charge state of a dot, we integrate this electron density over the dot area $A_{dot}$ (defined by a plunger gate), then take the floor of the result:
$
n_{dot} = \Big\lfloor \iint_{A_{dot}} n_{2D} \, dA_{dot} \Big\rfloor
$. 
In practice, this integral is approximated numerically. A naive approach—summing the values in the corresponding subarray and multiplying by the area per pixel—tends to significantly overestimate the electron count. To improve accuracy, we first up-sample the $n_{2D}$ values in the dot region using cubic spline interpolation (order 3)~\cite{de1972calculating}, doubling the resolution. We then perform the integration using Simpson’s rule along both axes.

To simulate the triple-point feature observed in charge stability diagrams, we applied a constant energy penalty for each electron already present in neighboring dots when adding an additional electron to the dot.


\subsection{Integration as a Simulated Backend}
\label{sec:integrating-as-a-simulated-backend}

\begin{figure*}[t]
  \centering
  \includegraphics[width=0.85\textwidth]{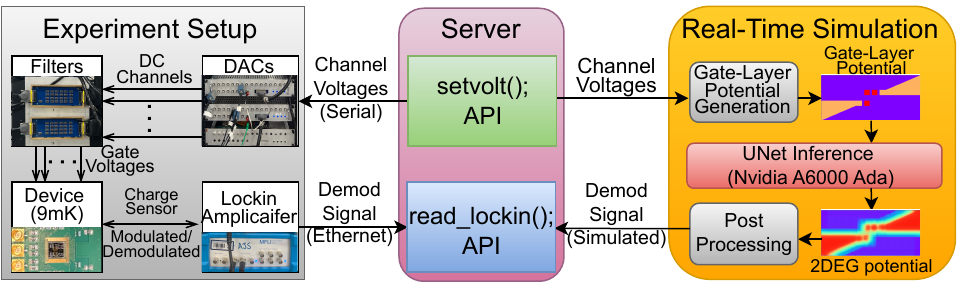}
  \caption{Experiment setup, server, and real-time simulation backend}\vspace{-5pt}
  \label{fig:simulation-architecture}
\end{figure*}

We build a complete real-time simulator for the 2-dot device by integrating the post-processing stage to the 2DEG potential output from the U-Net. To integrate the simulator into our experiment software infrastructure, we implement the simulated versions of the \textbf{set} and \textbf{read} operations described in Section~\ref{sec:background} used in experiments.

In Figure~\ref{fig:simulation-architecture}, the boxes "Experiment Setup" and "Server" illustrate the hardware and software stacks involved in device operation. On the software side, a server coordinates the experimental operations. It exposes two key APIs: \texttt{setvolt}, which sets gate voltages and corresponds to \textbf{set}, and \texttt{read\_lockin}, which reads the demodulated signal from the charge sensor and corresponds to \textbf{read}.

In a physical experiment, \texttt{setvolt} sets the voltage on the digital-to-analog converter (DAC) channel corresponding to a gate. The DAC output passes through a filter chain before reaching the gate on the device located in the 9~mK chamber of a dilution refrigerator. Meanwhile, \texttt{read\_lockin} operates a lock-in amplifier through an Ethernet connection and retrieves the demodulated signal, which reflects the conductance of the charge sensor.

For the simulated experiment, the "Server" and "Real-Time Simulation" boxes in Figure~\ref{fig:simulation-architecture} illustrate how the APIs are implemented for simulation. In the simulated \texttt{setvolt}, the gate-voltage pair is passed directly to the simulator rather than to a DAC. The simulator constructs the gate-layer potential by assigning the specified voltage to the corresponding gate region, then runs the U-Net model on an NVIDIA A6000 Ada GPU to generate the corresponding 2DEG potential. This potential is post-processed using the charge state model described earlier to produce a simulated demodulated signal. When the simulated \texttt{read\_lockin} API is called, the simulator returns the most recent simulated demodulated signal.

%% file: tex/7-Evaluation.tex
\section{Evaluation}
\label{sec:evaluation}

In this section, we evaluate NeuroQD by 1) comparing the accuracy and performance of the trained U-Net model to the COMSOL simulation for 2DEG potential inference, and 2) Compare the latency and simulated behavior of the simulated backend in Figure~\ref{fig:simulation-architecture} to real devices.

\subsection{Comparing U-Net to COMSOL simulation}
\label{sec:comparing-U-Net-to-comsol-simulation}
We construct a comprehensive set of test samples on six near-term devices ranging from 2 to 12 dots to benchmark the trained U-Net model against the COMSOL baseline. For accuracy comparison, we use the 2DEG computed by COMSOL as the ground truth and evaluate the percentage accuracy of the 2DEG computed by U-Net. For performance comparison, we measure the total runtime for computing the 2DEG for the test samples. We further extend the device size to 99 dot to evaluate the performance and accuracy trend for scalability.

\subsubsection{Devices}
The six near-term dot array devices are shown in the left column of Figure~\ref{fig:comsol-devices} in the appendix. Their geometries are defined by fixing the gate dimensions and heterostructures and by extending/shrinking the number of dots. Each of these 6 devices corresponds to a device architecture previously fabricated and used in experiments~\cite{zajac2018resonantly, borjans2021spin, philips2022universal, mills2019shuttling, george202412}. 


 The extended set of devices for trend analysis incrementally adds 3 dots per device, following the same heuristic of extending 6qd $\to$ 9qd $\to$ 12qd. The right column of Figure~\ref{fig:comsol-devices} in the appendix shows the extended set of devices. 


\subsubsection{Test Samples}
\label{sec:test-samples}

\begin{table}[b]
 \caption{Test set sample decomposition}
 \vspace{-5pt}
    \resizebox{0.95\columnwidth}{!}{
        \begin{tabular}{|c|c|c|c|c|c|c|c|}
            \hline
            & \multirow{2}{*}{All-Gate} & \multirow{2}{*}{Dot-Array} & \multicolumn{4}{c|}{Sensor-Dot(s)} & \multirow{2}{*}{Turn-On}\\
            \cline{4-7}
            & & & SD1 & SD2 & SD3 & SD4 & \\
            \hline
            2qd & 2500 & 1440 & 960 & N.A. & N.A. & N.A. & 100 \\
            \hline
            3qd & 2500 & 1600 & 800 & N.A. & N.A. & N.A. & 100 \\
            \hline
            4qd & 2500 & 1333 & 533 & 533 & N.A. & N.A. & 100\\
            \hline
            6qd & 2500 & 1527 & 436 & 436 & N.A. & N.A. & 100 \\
            \hline
            9qd & 2500 & 1500 & 300 & 300 & 300 & N.A. & 100\\
            \hline
            12qd & 2500 & 1485 & 228 & 228 & 228 & 228 & 100 \\
            \hline
        \end{tabular}
    }
   
    \label{test-set-sample-decomposition}
\end{table}

We generate 5000 test samples on each near-term device. Table \ref{test-set-sample-decomposition} shows the four types of samples included in the 5000 samples for each device. The all-gate samples, which make up half of the dataset, are generated using LHS over the full gate voltage space. The remaining three types of sample reflect gate voltage configurations applied in experiments.
\begin{itemize}
    \item Turn-on samples consist of evenly spaced voltages in the 0–1V range applied to all gates. These correspond to sweeping voltages from 0 to 1V on all gates when measuring the turn-on curve of a device.
    \item Sensor-dot samples are generated using LHS within the subspace of each individual charge sensor (sensor dot). These configurations are used during the sequential tuning of the sensor dots. Since the number of sensor dots varies by device, entries for nonexistent sensors (e.g. SD3 and SD4 on 4qd, which only has two sensors) are marked as N.A.
    \item Dot-array samples involve LHS within the dot-array subspace and represent configurations used to tune the electron occupancy of the qubit dots.
\end{itemize}
The number of samples in each type of sample is proportional to the number of gates involved. 


For trend analysis, we generated 100 random samples on each device, including extended devices. These samples are used to benchmark and visualize the accuracy and performance trend.

\subsubsection{Running Test Samples in COMSOL}
\label{sec:running-test-samples-in-comsol}
We run the test samples in COMSOL to generate the ground-truth 2DEG potentials and measure the runtimes. The measured runtime includes only the steps relevant to using COMSOL as a simulator: specifically, the 'Apply Voltages' and 'Run Solver' stages illustrated in Figure \ref{fig:overview}. All benchmarks are obtained using COMSOL 6.0 running on Windows 10 with a Ryzen Threadripper 3960X processor with the workload parallelized with 48 threads. The detailed results of running the test samples in COMSOL are shown in Table~\ref{COMSOL-test-samples} in the appendix.

\subsubsection{Running Test Samples on U-Net Model}
\label{sec:running-test-samples-on-U-Net-model}
We benchmark the accuracy of the U-Net model using the converged test samples from COMSOL. Accuracy is evaluated using the symmetric mean absolute percentage error (sMAPE), defined as:
$
\text{sMAPE} = \frac{1}{n} \sum_{i=1}^{n} \frac{|A_i - P_i|}{(|A_i| + |P_i|)/2 + \epsilon}, \
\text{Accuracy} = 1 - \frac{\text{sMAPE}}{200\%}
$
where $A_i$ and $P_i$ are the ground-truth and predicted 2DEG potential values at point $i$, respectively, and $n$ is the total number of points in the 2DEG potential. An $\epsilon = 10^{-10}$ is introduced in the denominator to avoid division by 0. Inference is performed using Python 3.13 and PyTorch 2.6 on an NVIDIA A6000 Ada GPU running under Windows 10. We run the augmented test sets through the U-Net and report the total number of augmented samples along with the average accuracy for each test set. These results are summarized in Table \ref{model-accuracies}. The detailed results of running the test samples on U-Net are shown in Table~\ref{model-accuracies} and Table~\ref{model-runtimes} in the Appendix.

\begin{figure}[t]
  \centering
  \includegraphics[width=\columnwidth]{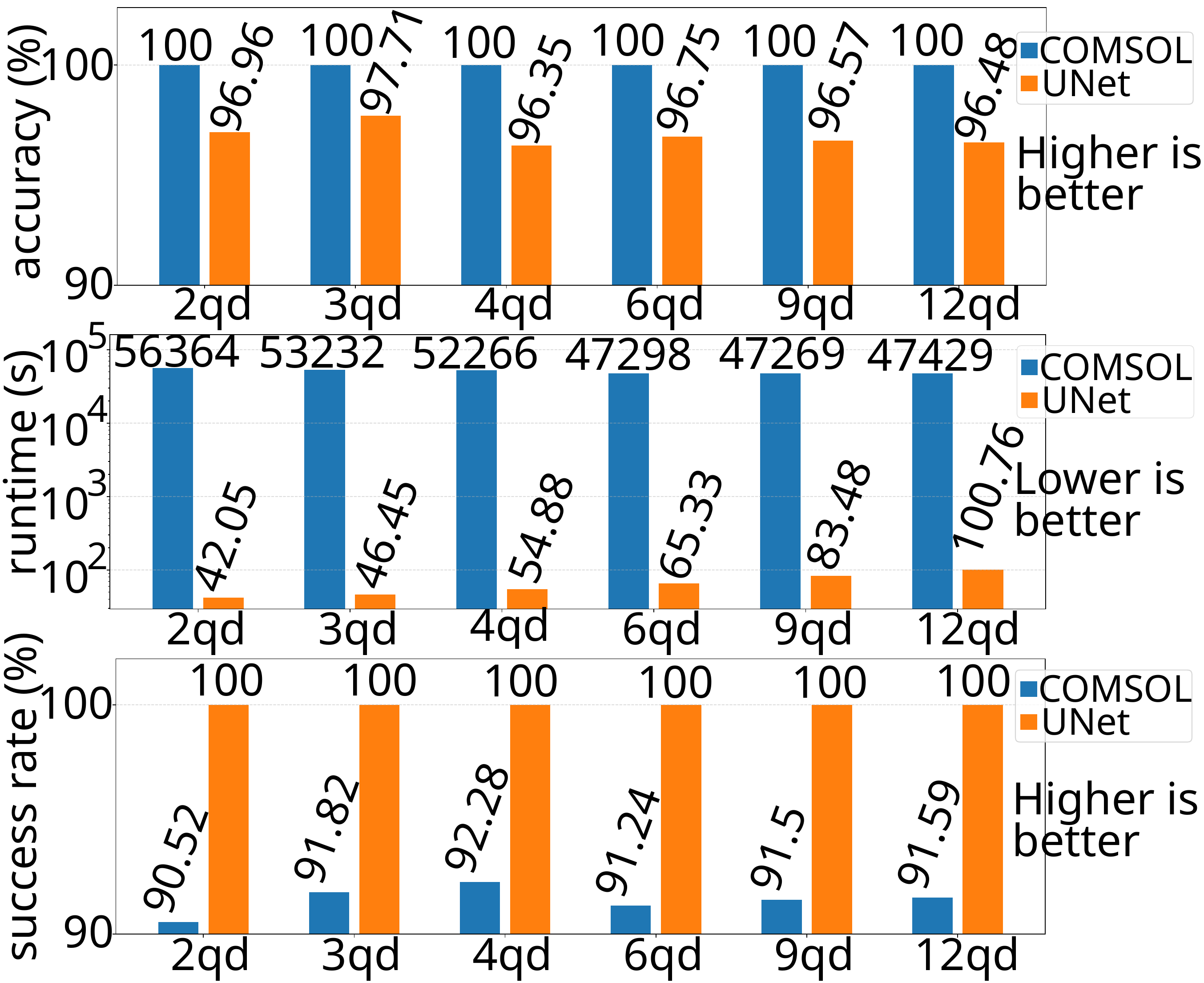}\vspace{-5pt}
  \caption{Comparison against the COMSOL baseline}\vspace{-5pt}
  \label{fig:comparison}
\end{figure}

\begin{figure}[t]
    \centering
    \includegraphics[width=\columnwidth]{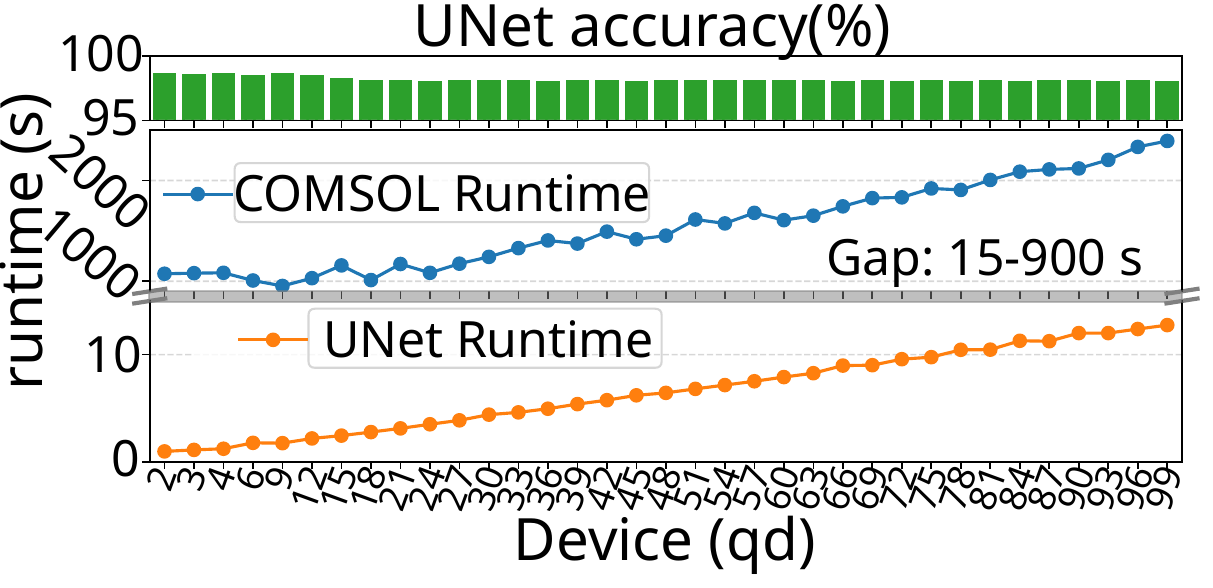}\vspace{-5pt}
    \caption{Accuracy and runtime trends}
    \label{fig:extra-compare}
\end{figure}
\vspace{-5pt}




Figure~\ref{fig:comparison} summarizes the overall accuracy, runtime, and success rate of COMSOL and the U-Net model on the six near-term devices in bar chart form. As shown, the U-Net model achieves 2–3 orders of magnitude speedup while achieving >96\% accuracy compared to the COMSOL baseline. In addition, it eliminates the convergence failure issue of COMSOL (100\% success rate compared to COMSOL's 91.49\%). 



\subsubsection{Trend Analysis}
\label{sec:trend-analysis-beyond-12qd}
The trend analysis follows the same procedures for accuracy and runtime benchmarking using the 100 samples generated for each device. Figure~\ref{fig:extra-compare} plots from top to bottom U-Net accuracy, COMSOL runtime, and U-Net runtime with shared horizontal axis representing the devices from 2qd to 99qd. The accuracy trend suggests that the trained model can generalize the potential mapping far beyond 12qd without significant drop in accuracy. The runtime trends for both COMSOL and U-Net are linear. The COMSOL runtime trend from 2qd to 24qd shows insufficient problem size for 48 thread parallelism (where parallelism overhead dominates the runtime) as suggested by the relatively flat or even decreasing trend on these smaller devices. This explains the strange runtime trend in Figure~\ref{fig:comparison} for near-term devices. For devices larger than 24qd, a steady linear trend is observed, suggesting full advantage of parallelism. The runtime trend of U-Net remains linear with less than 15 seconds runtime for the largest 99qd, suggesting a robust performance advantage as device scales.

\subsection{Comparing to Real Devices}
\label{sec:comparing-the-full-simulator-to-real-devices}
We conduct four experiments on our simulated backend, modeling a 2qd device. These experiments span from a cold start to a fully tuned charge state. We then compare the resulting simulated plots and runtime with corresponding measurements from two real devices operated under the same control software setup.

\subsubsection{Devices}
\label{sec:devices}

The experimental data used for comparison were collected from two devices, their dose test images are shown in Figure~\ref{fig:dose} in the appendix.

The first device (referred to as Device1) is a 2-dot shuttling device (top row), with gate dimensions matching the simulated 2qd device. The second device is a 4-dot device but operated as a 2-dot device using the left 2 dots, also with gate dimensions matching the simulated 2qd device. fabricated using the same layout and dimensions captured in Figure~\ref{fig:device-geometry}.

The gate features included/not included in the simulated backend are clearly indicated by the red rectangles (simulated region) and cyan rectangles (not included in simulation) in Figure~\ref{fig:dose}. The results from device 1 were previously collected without runtime measurement, and the results from device 2 were collected from recent cool-down with runtime measurement. In comparing the simulated plots to the experimentally obtained plots, we focus on key features and physical quantities. We also compare the simulation runtime to the runtime measured from the most recent cool-down.

\subsubsection{Turn-on}
\label{sec:turn-on}
\begin{figure}[h]
  \centering
  \includegraphics[width=\columnwidth]{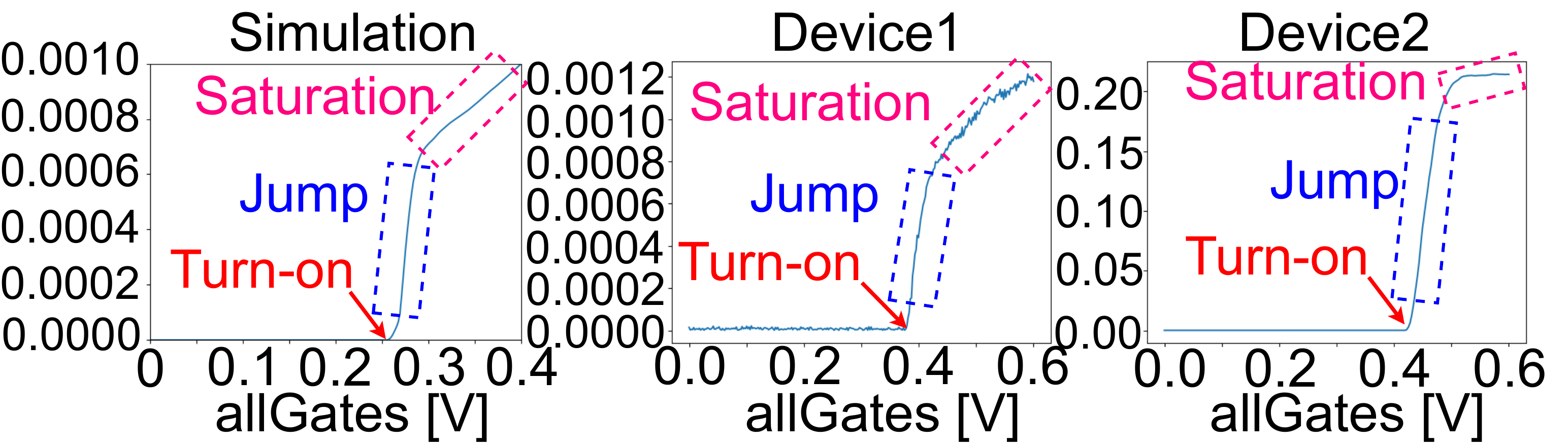}
  \caption{Simulated turn-on curve}\vspace{-5pt}
  \label{fig:simulated-turn-on}
\end{figure}

The first experiment is the turn-on experiment, where we sweep the voltages on all gates until the conduction saturates. The key features of this experiment are a sudden "jump" in conduction when the "allGate" voltage reaches the turn-on threshold, followed by a slower rise in conduction as it enters the saturated region. As shown in Figure~\ref{fig:simulated-turn-on}, these features were observed on both devices and simulated by the simulation.

\subsubsection{Slit vs Sensor Sweep}
\label{sec:slit-vs-sensor-sweep}
\begin{figure}[h]
  \centering
  \includegraphics[width=\columnwidth]{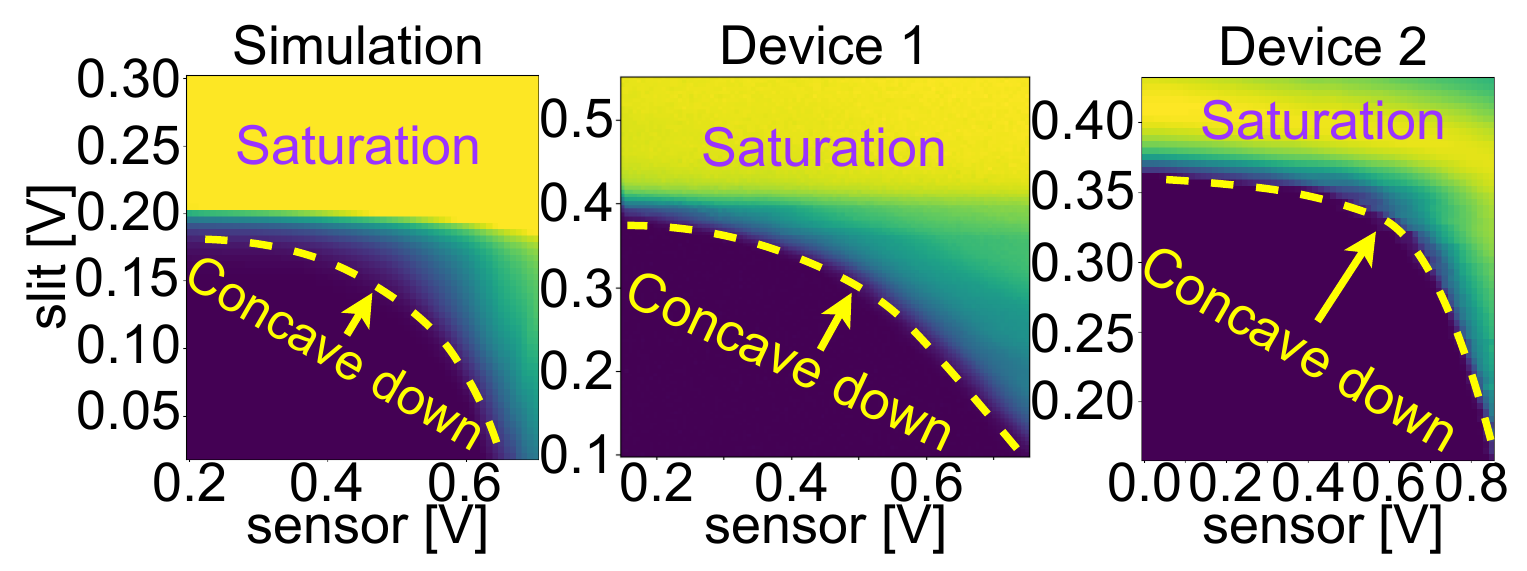}
  \caption{Simulated slit vs sensor sweep}\vspace{-5pt}
  \label{fig:simulated-slit-vs-sensor}
\end{figure}

The second experiment is the slit gate vs sensor sweep experiment, where we do a 2d sweep of the voltages on the sensor slit gate (S3 gate referring back to Figure~\ref{fig:gate-layer-potential}) and the voltages on the three sensor finger gates (B4, P3, P5 referring back to Figure~\ref{fig:gate-layer-potential}). The first key feature in this experiment is the saturation region on top of the plot caused by the slit gate voltage turning on the conduction in its gate region. Outside the saturation region, the conduction caused by the sensor gates should form a lighter region that concaves down. As shown in Figure~\ref{fig:simulated-slit-vs-sensor}, these features were observed in both devices and simulated by simulation.

\subsubsection{Coulomb Peaks}
\label{sec:coulomb-peaks}
\begin{figure}[h]
  \centering
  \includegraphics[width=\columnwidth]{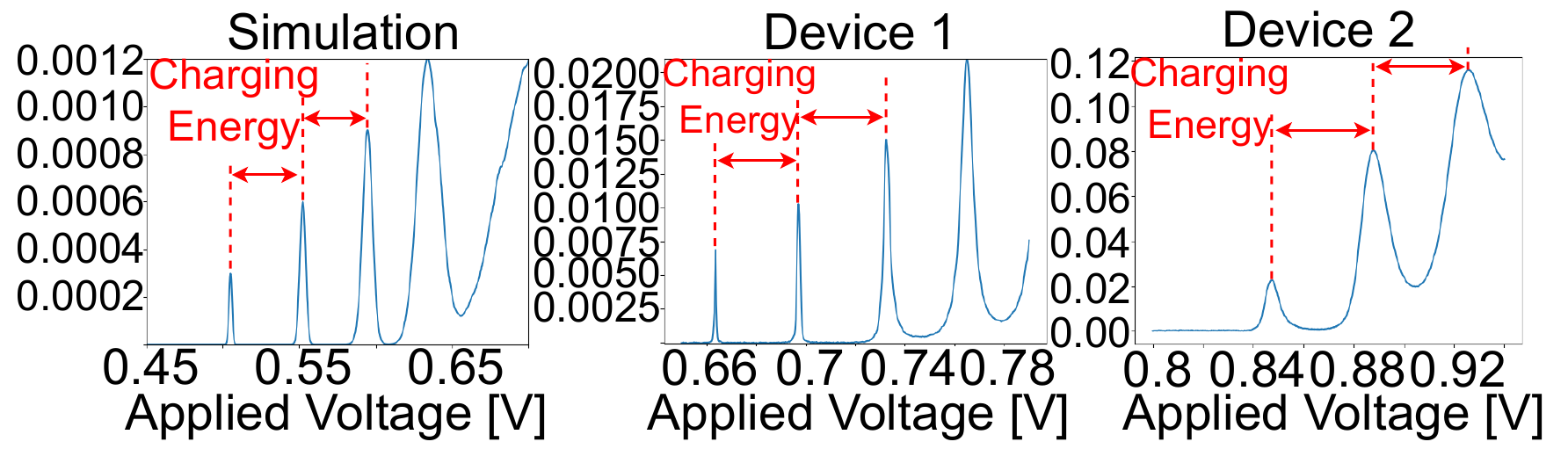}
  \caption{Simulated Coulomb peaks}\vspace{-5pt}
  \label{fig:coulomb-peaks}
\end{figure}

The third experiment is the Coulomb Peaks experiment, where we fix the sensor slit voltage (below the level where it reaches saturation) and sweep the voltage on the sensor plunger gate (P3 in Figure~\ref{fig:gate-layer-potential}). The key feature is the discrete conduction caused by the Coulomb blockade effect. A key physical quantity that we can compare here is the charging energy, which is the distance between the conduction peaks. As shown in Figrue~\ref{fig:coulomb-peaks}, the conduction peaks were observed on both devices and present in the simulation. The simulated charging energies are around 0.05V, which is comparable to the 0.04V observed on both devices.

\subsubsection{Charge Stability Diagram}
\label{sec:charge-stability-diagram}
\begin{figure}[h]
  \centering
  \includegraphics[width=\columnwidth]{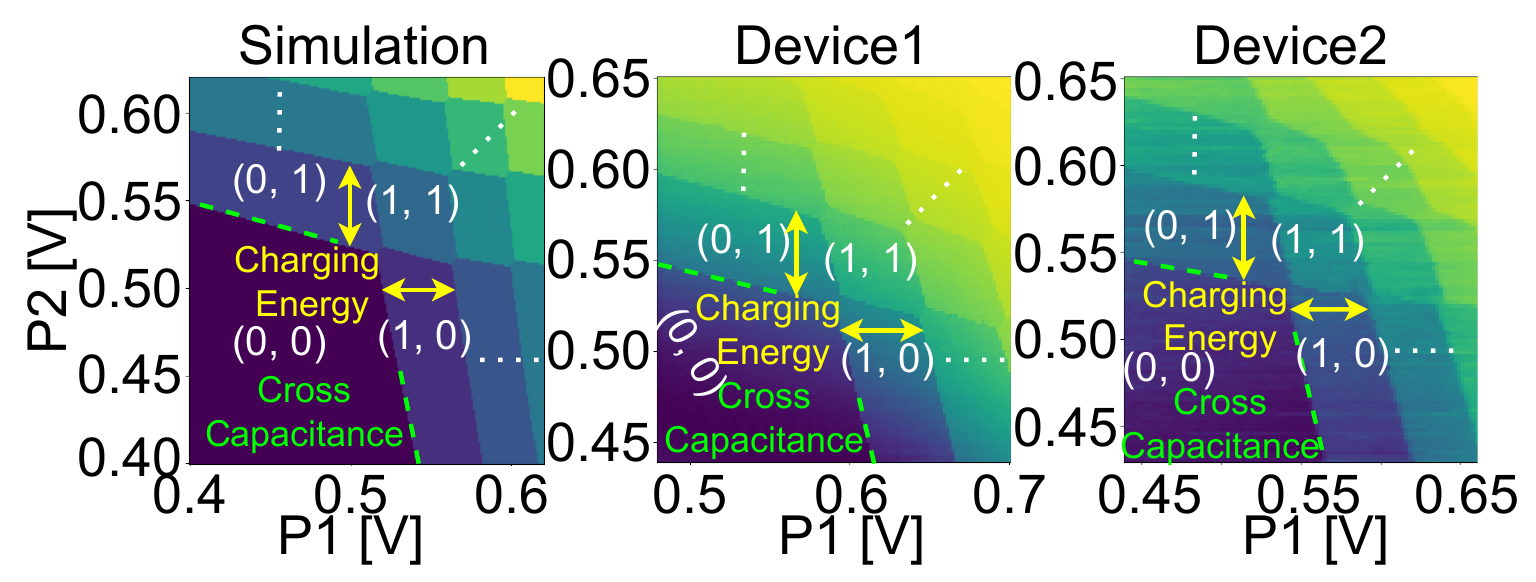}
  \caption{Simulated charge stability diagram}\vspace{-5pt}
  \label{fig:simulated-csd}
\end{figure}

The fourth experiment is the charge stability diagram (CSD) experiment, where we sweep the voltages on the qubit-dot plunger gates (P1 and P2 in Figure~\ref{fig:gate-layer-potential}). The key feature is the defined charge state regions in the P1, P2 gate voltage space. The tuples in Figure~\ref{fig:simulated-csd} indicate the number of electrons trapped in the dot under P1 and P2. Two key physical quantities that we can compare in this experiment are the charging energy, as defined by the spacing between the transition lines (yellow arrows in Figure~\ref{fig:simulated-csd}), and the cross-capacitance, as defined by the slopes of the transition lines (green lines in Figure~\ref{fig:simulated-csd}). The defined charge state regions are clearly visible in the simulated CSD, matching the shapes observed in Devices 1 and 2. The charging energy in the simulation (slightly less than 0.05V) is slightly less than that observed in device 1 and 2 (about 0.05V). The slopes of the transition lines (cross-capacitance) from the simulation are in good agreement with device 1 and device 2.

\subsubsection{Runtime Comparison}
\label{sec:runtime-comparison}
\begin{figure}[h]
  \centering
  \includegraphics[width=\columnwidth]{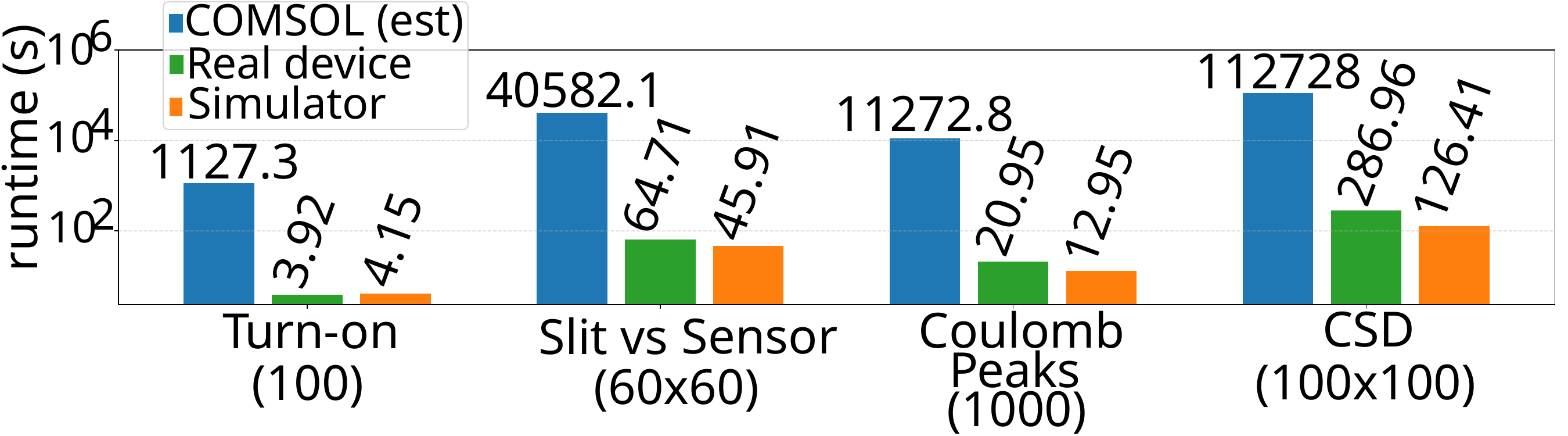}
  \caption{Runtime comparison}
  \label{fig:runtime-compare}
\end{figure}

We compare the simulation runtime with the runtime of device 2 in the four experiments to demonstrate real-time performance comparable to live device measurement. For each experiment, the runtime for device 2 was measured by minimizing the wait time without compromising the quality of the feature in the plot. Figure~\ref{fig:runtime-compare} shows the measured runtimes of running the four experiments using the simulator, on device 2 (labeled "Real device"), and using COMSOL simulation. The COMSOL runtime is estimated by multiplying the averaged COMSOL simulation time for one configuration by the number of voltage configurations required in the corresponding experiment. Overall, the runtimes of our simulator can keep up with live measurements and are orders of magnitude faster than the estimated runtimes of COMSOL.

%% file: tex/Appendix.tex
\section{Data augmentation}
\label{sec:data-augmentation}

We augment the training data set by applying two preprocessing steps to the raw potential profiles generated by COMSOL: gate-layer potential approximation and data augmentation via rotation and reflection.

\begin{figure}[H]
  \centering
  \includegraphics[width=\columnwidth]{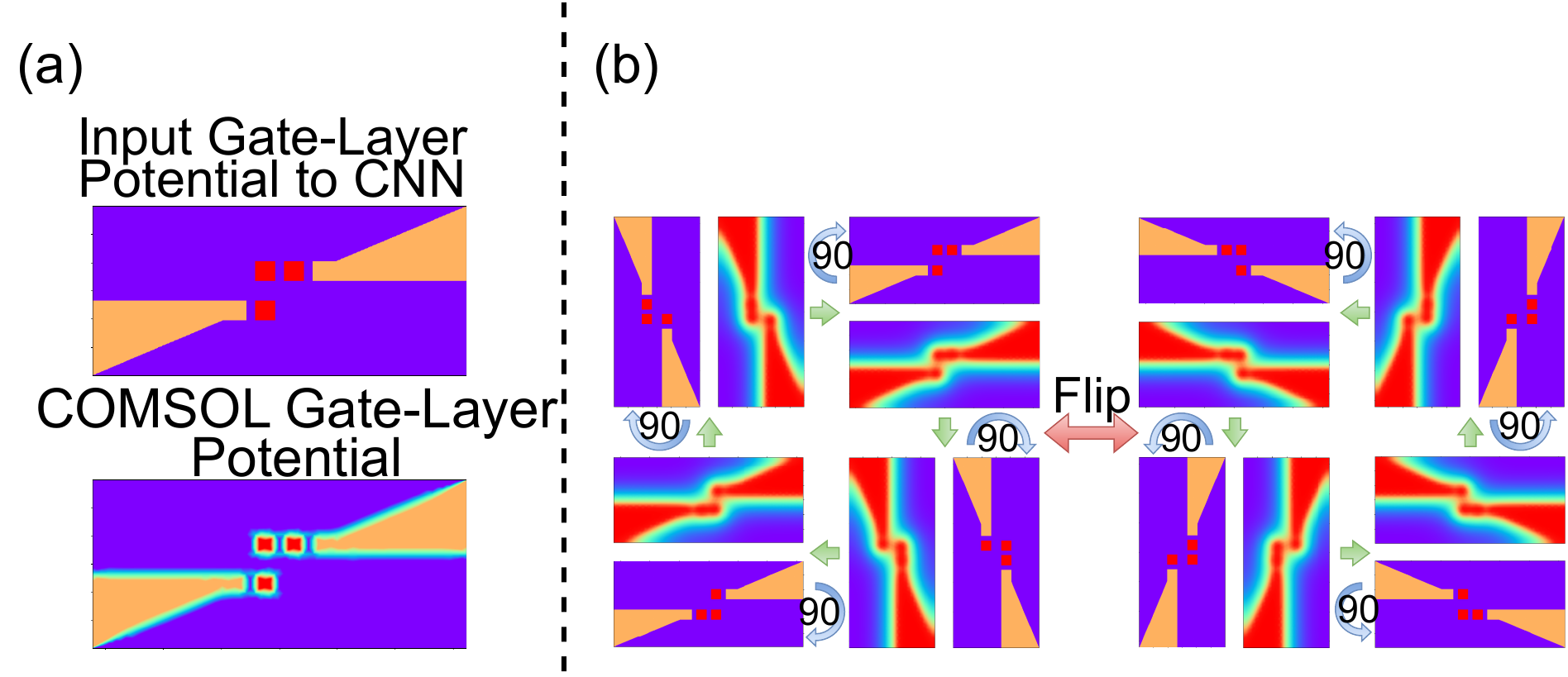}
  \caption{(a) Gate-layer potential approximation, (b) Training data augmentation via rotation and reflection}
  \label{fig:data-augmentation}
\end{figure}

\textbf{Gate-Layer Potential Approximation} As shown on the top left in Figure~\ref{fig:data-augmentation} (a), the gate-layer potential is defined by assigning voltages to specific gate regions. However, the actual gate-layer potential in COMSOL has subtle features along the edges of these regions, where the potential slightly drops, forming thin border bands. For our purposes, we do not use COMSOL gate-layer potentials for training and generate training inputs by assigning voltages directly to gate regions, which disregards these subtle features.

\textbf{Data Augmentation via Rotation and Reflection} We apply geometric transformations to the input/output pairs. The target transformation is invariant under rotation and reflection. \textbf{Rotations:} Three additional samples are generated by rotating both the input and output by 90°, 180°, and 270° (Figure~\ref{fig:data-augmentation} (b), left).
\textbf{Reflections:} Four additional samples are generated by flipping the original sample horizontally and again apply all three rotations to the flipped version (Figure~\ref{fig:data-augmentation} (b), right). In total, each original input/output pair yields seven new samples, increasing the dataset size by a factor of 8.

\section{Devices for benchmarking}

\begin{figure}[H]
  \includegraphics[width=\columnwidth]{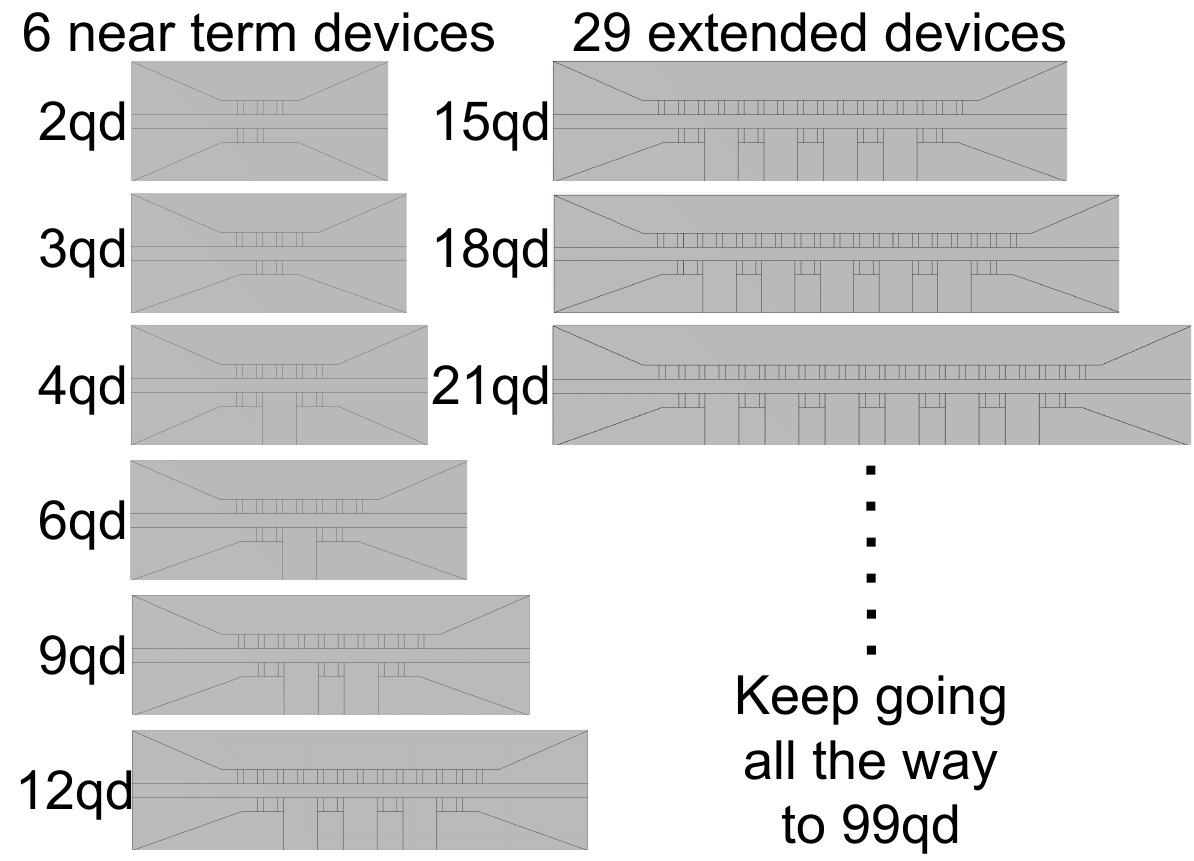}
  \caption{COMSOL devices}
  \label{fig:comsol-devices}
\end{figure}

\begin{figure}[H]
  \includegraphics[width=\columnwidth]{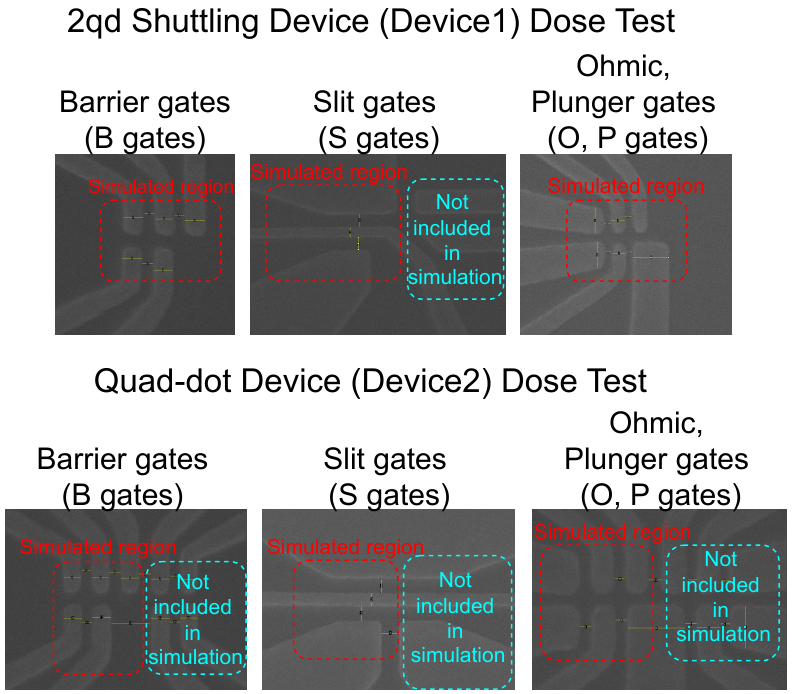}
  \caption{Device 1 and device 2 dose test}
  \label{fig:dose}
\end{figure}

Figure~\ref{fig:comsol-devices} summarizes the layout of the 35 devices (6 near-term devices and 29 extended devices) used in the experiments in Section~\ref{sec:comparing-U-Net-to-comsol-simulation}. Figure~\ref{fig:dose} shows the dose test images of the two real devices used in Section~\ref{sec:comparing-the-full-simulator-to-real-devices} with simulated/operated regions highlighted in red. 

\section{U-Net architecture}

Table~\ref{unet-summary} summarizes the architecture of the U-Net model used in this paper. The model was trained with the mean squared error loss function for 30 epochs.

\begin{table*}[t]
\centering
\caption{U-Net architecture summary}
\resizebox{\textwidth}{!}{
\begin{tabular}{|l|l|c|c|c|c|c|}
\hline
\textbf{Layer} & \textbf{Operation} & \textbf{Channels In} & \textbf{Channels Out} & \textbf{Kernel Size} & \textbf{Stride} & \textbf{Output Shape} \\
\hline
\multirow{2}{*}{\textbf{Encoder 1}} & Conv2D$\times$2 + ReLU + BN & 1 & 64 & 3$\times$3 & 1 & H$\times$W$\times$64 \\
\cline{2-7}
 & MaxPool2D & 64 & 64 & 2$\times$2 & 2 & H/2$\times$W/2$\times$64 \\
\hline
\multirow{2}{*}{\textbf{Encoder 2}} & Conv2D$\times$2 + ReLU + BN & 64 & 128 & 3$\times$3 & 1 & H/2$\times$W/2$\times$128 \\
\cline{2-7}
 & MaxPool2D & 128 & 128 & 2$\times$2 & 2 & H/4$\times$W/4$\times$128 \\
\hline
\multirow{2}{*}{\textbf{Encoder 3}} & Conv2D$\times$2 + ReLU + BN & 128 & 256 & 3$\times$3 & 1 & H/4$\times$W/4$\times$256 \\
\cline{2-7}
 & MaxPool2D & 256 & 256 & 2$\times$2 & 2 & H/8$\times$W/8$\times$256 \\
\hline
\textbf{Bottleneck} & Conv2D$\times$2 + ReLU + BN & 256 & 512 & 3$\times$3 & 1 & H/8$\times$W/8$\times$512 \\
\hline
\multirow{3}{*}{\textbf{Decoder 3}} & ConvTranspose2D & 512 & 256 & 2$\times$2 & 2 & H/4$\times$W/4$\times$256 \\
\cline{2-7}
 & Concatenate & 256+256 & 512 & - & - & H/4$\times$W/4$\times$512 \\
\cline{2-7}
 & Conv2D$\times$2 + ReLU + BN & 512 & 256 & 3$\times$3 & 1 & H/4$\times$W/4$\times$256 \\
\hline
\multirow{3}{*}{\textbf{Decoder 2}} & ConvTranspose2D & 256 & 128 & 2$\times$2 & 2 & H/2$\times$W/2$\times$128 \\
\cline{2-7}
 & Concatenate & 128+128 & 256 & - & - & H/2$\times$W/2$\times$256 \\
\cline{2-7}
 & Conv2D$\times$2 + ReLU + BN & 256 & 128 & 3$\times$3 & 1 & H/2$\times$W/2$\times$128 \\
\hline
\multirow{3}{*}{\textbf{Decoder 1}} & ConvTranspose2D & 128 & 64 & 2$\times$2 & 2 & H$\times$W$\times$64 \\
\cline{2-7}
 & Concatenate & 64+64 & 128 & - & - & H$\times$W$\times$128 \\
\cline{2-7}
 & Conv2D$\times$2 + ReLU + BN & 128 & 64 & 3$\times$3 & 1 & H$\times$W$\times$64 \\
\hline
\textbf{Output} & Conv2D & 64 & 1 & 1$\times$1 & 1 & H$\times$W$\times$1 \\
\hline
\end{tabular}
}

\label{unet-summary}
\end{table*}

\section{Detailed Result Tables}
Table~\ref{COMSOL-test-samples} summarizes the results of running test samples in COMSOL. The “\#converged” row indicates the number of samples for which COMSOL successfully computed the 2DEG potential. Samples that failed to converge represent unsuccessful attempts where COMSOL could not produce a solution. The “success\%” reflects the ratio of converged samples to the total. Runtime measurements are also provided for each set of samples. Table~\ref{model-accuracies} summarizes the accuracies on running test samples on the trained U-Net model. Rotational augmentation was applied to all converged samples—following the same procedure shown in the left half of Figure \ref{fig:data-augmentation}—for more comprehensive accuracy benchmarking. Reflectional augmentation is not applied because the gate layout of larger devices are symmetric. Table~\ref{model-runtimes} summarizes the runtimes of running unaugmented test samples on the trained U-Net model.

\begin{table*}[t]
\caption{Results of running test samples in COMSOL}
    \resizebox{\textwidth}{!}{
        \begin{tabular}{|c|c|c|c|c|c|c|c|c|c|}
            \hline
            & & \multirow{2}{*}{All-Gate} & \multirow{2}{*}{Dot-Array} & \multicolumn{4}{c|}{Sensor-Dot(s)} & \multirow{2}{*}{Turn-On} & \multirow{2}{*}{Overall}\\
            \cline{5-8}
            & & & & SD1 & SD2 & SD3 & SD4 & & \\
            \hline
            \multirow{4}{*}{2qd} & \#samples & 2500 & 1440 & 960 & N.A. & N.A. & N.A. & 100 & 5000\\
                                 \cline{2-10}
                                 & \#converged & 2148 & 1368 & 913 & N.A. & N.A. & N.A. & 97 & 4526 \\
                                 \cline{2-10}
                                 & success\% & 85.92\% & 95\% & 95.1\% & N.A. & N.A. & N.A. & 97\% & 90.52\%\\
                                 \cline{2-10}
                                 & runtime & 28895s & 16936s & 9790s & N.A. & N.A. & N.A. & 743s & 56364s\\
            \hline
            \multirow{4}{*}{3qd} & \#samples & 2500 & 1600 & 800 & N.A. & N.A. & N.A. & 100 & 5000 \\
                                 \cline{2-10}
                                 & \#converged & 2159 & 1565 & 770 & N.A. & N.A. & N.A. & 97 & 4591\\
                                 \cline{2-10}
                                 & success\% & 83.36\% & 97.81\% & 96.25\% & N.A. & N.A. & N.A. & 97\% & 91.82\%\\
                                 \cline{2-10}
                                 & runtime & 27893s & 17612s & 7018s & N.A. & N.A. & N.A. & 707s & 53232s\\
            \hline
            \multirow{4}{*}{4qd} & \#samples & 2500 & 1333 & 533 & 533 & N.A. & N.A. & 100 & 4999\\
                                 \cline{2-10}
                                 & \#converged & 2140 & 1311 & 533 & 533 & N.A. & N.A. & 96 & 4613\\
                                 \cline{2-10}
                                 & success\% & 85.6\% & 98.35\% & 100\% & 100\% & N.A. & N.A. & 96\% & 92.28\%\\
                                 \cline{2-10}
                                 & runtime & 28119s & 13493s & 4956 & 4990s & N.A. & N.A. & 708s & 52266s\\
            \hline
            \multirow{4}{*}{6qd} & \#samples & 2500 & 1527 & 436 & 436 & N.A. & N.A. & 100 & 4999 \\
                                 \cline{2-10}
                                 & \#converged & 2128 & 1478 & 427 & 431 & N.A. & N.A. & 97 & 4561\\
                                 \cline{2-10}
                                 & success\% & 85.12\% & 96.79\% & 97.94\% & 98.85\% & N.A. & N.A. & 97\% & 91.24\% \\
                                 \cline{2-10}
                                 & runtime & 25384s & 14332s & 3444s & 3506s & N.A. & N.A. & 632s & 47298s\\
            \hline
            \multirow{4}{*}{9qd} & \#samples & 2500 & 1500 & 300 & 300 & 300 & N.A. & 100 & 5000\\
                                 \cline{2-10}
                                 & \#converged & 2122 & 1456 & 300 & 300 & 300 & N.A. & 97 & 4575\\
                                 \cline{2-10}
                                 & success\% & 84.88\% & 97.07\% & 100\% & 100\% & 100\% & N.A. & 97\% & 91.5\%\\
                                 \cline{2-10}
                                 & runtime & 25067s & 14183 &  2398s & 2499s & 2473s & N.A. & 649s & 47269s\\
            \hline
            \multirow{4}{*}{12qd} & \#samples & 2500 & 1485 & 228 & 228 & 228 & 228 & 100 & 4997 \\
                                 \cline{2-10}
                                 & \#converged & 2135 & 1432 & 228 & 228 & 228 & 228 & 98 & 4577 \\
                                 \cline{2-10}
                                 & success\% & 85.4\% & 96.43\% & 100\% & 100\% & 100\% & 100\% & 98\% & 91.59\%\\
                                 \cline{2-10}
                                 & runtime & 26158s & 13965s & 1824s & 1902s & 1090s & 1842s & 648s & 47429s \\
            \hline
        \end{tabular}
    }
    
    \label{COMSOL-test-samples}
\end{table*}

\begin{table*}[t]
 \caption{UNet Model Accuracies on converged test samples with rotational augmentation}
    \resizebox{\textwidth}{!}{
        \begin{tabular}{|c|c|c|c|c|c|c|c|c|c|}
            \hline
            & & \multirow{2}{*}{All-Gate} & \multirow{2}{*}{Dot-Array} & \multicolumn{4}{c|}{Sensor-Dot(s)} & \multirow{2}{*}{Turn-On} & \multirow{2}{*}{Overall}\\
            \cline{5-8}
            & & & & SD1 & SD2 & SD3 & SD4 & & \\
            \hline
            \multirow{2}{*}{2qd} & \#samples & 8592 & 5472 & 3652 & N.A. & N.A. & N.A. & 388 & 18104\\
                                 \cline{2-10}
                                 & accuracy & 98.69\% & 97.2\% & 91.44\% & N.A. & N.A. & N.A. & 94.71\% & 96.96\%\\
            \hline
            \multirow{2}{*}{3qd} & \#samples & 8636 & 6260 & 3080 & N.A. & N.A. & N.A. & 388 & 18364\\
                                 \cline{2-10}
                                 & accuracy  & 98.63\% & 98.97\% & 90.3\% & N.A. & N.A. & N.A. & 94.86\% & 97.71\%\\
            \hline
            \multirow{2}{*}{4qd} & \#samples & 8560 & 5244 & 2132 & 2132 & N.A. & N.A. & 384 & 18452\\
                                 \cline{2-10}
                                 & accuracy  & 98.67\% & 93.97\% & 88.61\% & 88.62\% & N.A. & N.A. & 94.89\% & 96.35\%\\
            \hline
            \multirow{2}{*}{6qd} & \#samples & 8512 & 5912 & 1708 & 1724 & N.A. & N.A. & 388 & 18244\\
                                 \cline{2-10}
                                 & accuracy  & 98.64\% & 98.95\% & 88.47\% & 88.48\% & N.A. & N.A. & 94.87\% & 96.75\%\\
            \hline
            \multirow{2}{*}{9qd} & \#samples & 8488 & 5824 & 1200 & 1200 & 1200 & N.A. & 388 & 18300\\
                                 \cline{2-10}
                                 & accuracy  & 98.65\% & 98.93\% & 88.41\% & 87.15\% & 88.42\% & N.A. & 94.95\% & 96.57\%\\
            \hline
            \multirow{2}{*}{12qd} & \#samples & 8540 & 5728 & 912 & 912 & 912 & 912 & 392 & 18308\\
                                 \cline{2-10}
                                 & accuracy  & 98.62\% & 98.93\% & 88.35\% & 87.25\% & 87.24\% & 88.36\% & 94.86\% & 96.48\%\\
            \hline
        \end{tabular}
    }
   
    \label{model-accuracies}
\end{table*}

\begin{table*}[t]
   \caption{UNet Model runtimes on the original test samples}\vspace{-5pt}
    \resizebox{\textwidth}{!}{
        \begin{tabular}{|c|c|c|c|c|c|c|c|c|c|}
            \hline
            & & \multirow{2}{*}{All-Gate} & \multirow{2}{*}{Dot-Array} & \multicolumn{4}{c|}{Sensor-Dot(s)} & \multirow{2}{*}{Turn-On} & \multirow{2}{*}{Overall}\\
            \cline{5-8}
            & & & & SD1 & SD2 & SD3 & SD4 & & \\
            \hline
            \multirow{2}{*}{2qd} & \#samples & 2500 & 1440 & 960 & N.A. & N.A. & N.A. & 100 & 5000\\
                                 \cline{2-10}
                                 & runtime & 21.06s & 12.07s & 8.07s & N.A. & N.A. & N.A. & 0.85s & 42.05s\\
            \hline
            \multirow{2}{*}{3qd} & \#samples & 2500 & 1600 & 800 & N.A. & N.A. & N.A. & 100 & 5000\\
                                 \cline{2-10}
                                 & runtime  & 23.21s & 14.94s & 7.45s & N.A. & N.A. & N.A. & 0.94s & 46.45s \\
            \hline
            \multirow{2}{*}{4qd} & \#samples & 2500 & 1333 & 533 & 533 & N.A. & N.A. & 100 & 4999\\
                                 \cline{2-10}
                                 & runtime & 27.56s & 14.42s & 5.93s & 5.9s & N.A. & N.A. & 1.07s & 54.88s \\
            \hline
            \multirow{2}{*}{6qd} & \#samples & 2500 & 1527 & 436 & 436 & N.A. & N.A. & 100 & 4999\\
                                 \cline{2-10}
                                 & runtime & 32.91s & 19.84s & 5.62s & 5.66s & N.A. & N.A. & 1.3s & 65.33s \\
            \hline
            \multirow{2}{*}{9qd} & \#samples & 2500 & 1500 & 300 & 300 & 300 & N.A. & 100 & 5000\\
                                 \cline{2-10}
                                 & runtime & 41.25s & 24.73s & 4.97s & 4.94s & 4.95s & N.A. & 2.64s & 83.48s \\
            \hline
            \multirow{2}{*}{12qd} & \#samples & 2500 & 1485 & 228 & 228 & 228 & 228 & 100 & 4997\\
                                 \cline{2-10}
                                 & runtime & 50.39s & 29.98s & 4.59s & 4.6s & 4.6s & 4.6s & 2s & 100.76s\\
            \hline
        \end{tabular}
    }
 
    \label{model-runtimes}
\end{table*}
